\documentclass[twocolumn]{aastex62}
\usepackage[utf8]{inputenc}
\usepackage{longtable}
\usepackage{graphicx}
\usepackage{amsmath,amssymb}
\usepackage{color}
\usepackage{units}
\usepackage{epstopdf}
\usepackage{hyperref}
\usepackage{multirow}
\usepackage{url}

\usepackage{subfigure}
\usepackage{rotating}

\newcommand{\beq}{\begin{equation}}
\newcommand{\eeq}{\end{equation}}
\newcommand{\bdm}{\begin{displaymath}}
\newcommand{\edm}{\end{displaymath}}

\definecolor{Gray}{gray}{0.9}
\definecolor{orange}{rgb}{0.9,0.5,0}

\graphicspath{{./plots/}}

\begin{document}

\title{Teamwork Makes the Dream Work: Optimizing Multi-Telescope Observations of Gravitational-Wave Counterparts}

\author[0000-0002-8262-2924]{Michael W. Coughlin}
\affil{Division of Physics, Mathematics, and Astronomy, California Institute of Technology, Pasadena, CA 91125, USA}
\author[0000-0002-7686-3334]{Sarah Antier}
\affil{APC, UMR 7164, 10 rue Alice Domon et Léonie Duquet, 75205 Paris, France}
\author{David Corre}
\affil{LAL, Univ Paris-Sud, CNRS/IN2P3, Orsay, France}
\author{Khalid Alqassimi}
\affil{American University of Sharjah, Physics Department, PO Box 26666, Sharjah, UAE}
\author{Shreya Anand}
\affil{Division of Physics, Mathematics, and Astronomy, California Institute of Technology, Pasadena, CA 91125, USA}
\author{Nelson Christensen}
\affil{Artemis, Universit\'e C\^ote d'Azur, Observatoire C\^ote d'Azur, CNRS, CS 34229, F-06304 Nice Cedex 4, France}
\affil{Carleton College, Northfield, MN 55057, USA}
\author[0000-0003-4263-2228]{David A. Coulter}
\affil{Department of Astronomy and Astrophysics, University of California, Santa Cruz, CA 95064, USA}
\author{Ryan J. Foley}
\affil{Department of Astronomy and Astrophysics, University of California, Santa Cruz, CA 95064, USA}
\author{Nidhal Guessoum}
\affil{American University of Sharjah, Physics Department, PO Box 26666, Sharjah, UAE}
\author{Timothy M. Mikulski}
\affil{Carleton College, Northfield, MN 55057, USA}
\author{Mouza Al Mualla}
\affil{American University of Sharjah, Physics Department, PO Box 26666, Sharjah, UAE}
\author{Draco Reed}
\affil{Department of Astronomy and Astrophysics, University of California, Santa Cruz, CA 95064, USA}
\author{Duo Tao}
\affil{Division of Physics, Mathematics, and Astronomy, California Institute of Technology, Pasadena, CA 91125, USA}

\begin{abstract}
The ever-increasing sensitivity of the network of gravitational-wave detectors has resulted in the accelerated rate of detections from compact binary coalescence systems in the third observing run of Advanced LIGO and Advanced Virgo. 
Not only has the event rate increased, but also the distances to which phenomena can be detected, leading to a rise in the required sky volume coverage to search for counterparts. Additionally, the improvement of the detectors has resulted in the discovery of more compact binary mergers involving neutron stars, revitalizing dedicated follow-up campaigns.
While significant effort has been made by the community to optimize single telescope observations, using both synoptic and galaxy-targeting methods, less effort has been paid to coordinated observations in a network.
This is becoming crucial, as the advent of gravitational-wave astronomy has garnered interest around the globe, resulting in abundant networks of telescopes available to search for counterparts.
In this paper, we extend some of the techniques developed for single telescopes to a telescope network.
We describe simple modifications to these algorithms and demonstrate them on existing network examples.
These algorithms are implemented in the open-source software \texttt{gwemopt}, used by some follow-up teams, for ease of use by the broader community.
\linebreak
\end{abstract}

\section{Introduction}

\begin{figure}[t]
 \includegraphics[width=3.5in]{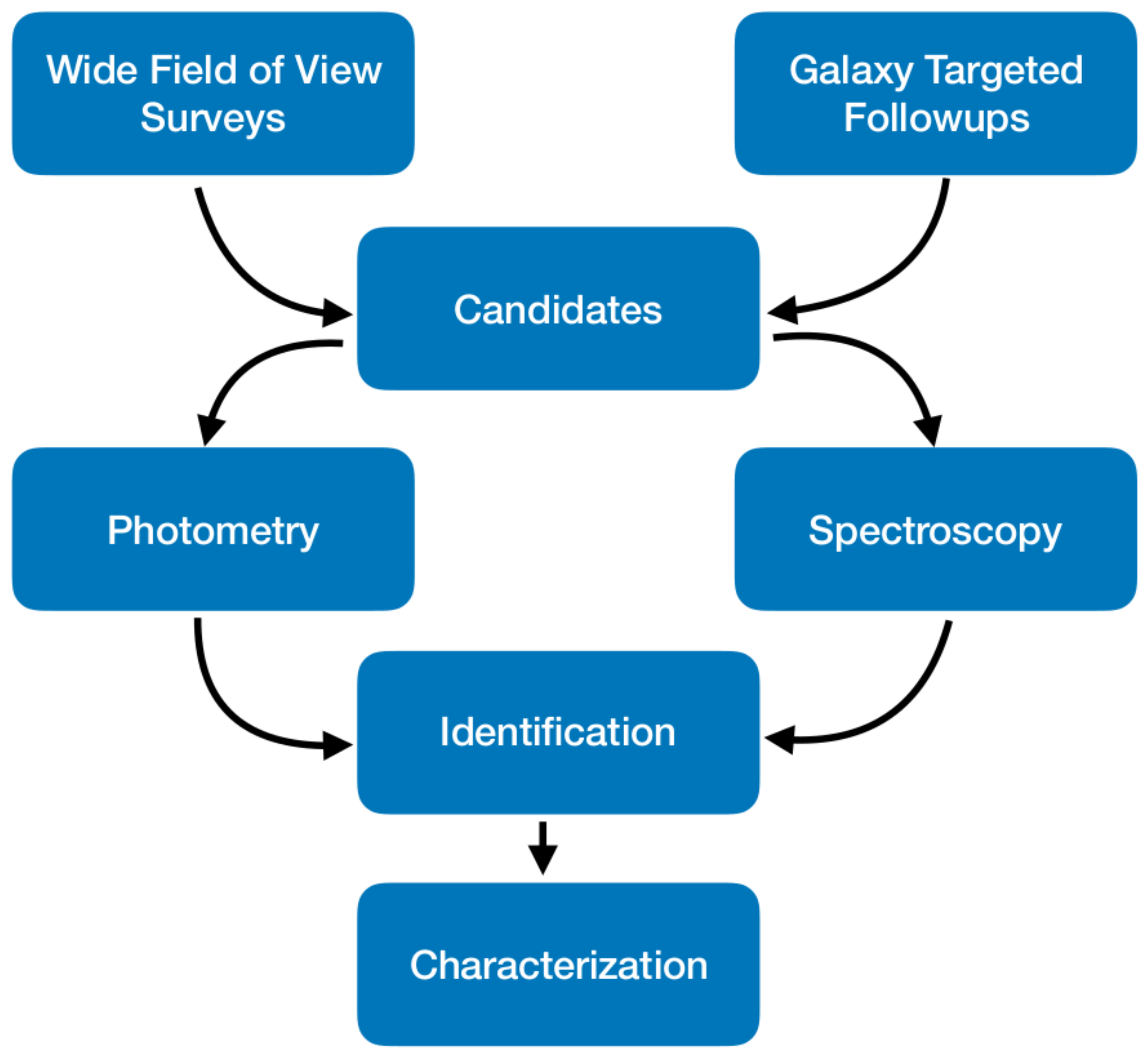}
  \caption{Flowchart of gravitational-wave electromagnetic counterpart follow-up strategy.}
 \label{fig:flowchart}
\end{figure} 

The science enabled by gravitational-wave astronomy is rapidly increasing as the sensitivity of the network of gravitational-wave detectors with Advanced LIGO \citep{aLIGO} and Advanced Virgo \citep{adVirgo} continues to improve.
The third observing run (O3), which began in April 2019, has already yielded the detection of many binary black hole systems \citep{SiEA2019,ShEA2019,ChEA2019,SiEA2019a,ChEA2019a,GhEA2019} and a few with at least one neutron star \citep{SiEA2019a,ChEA2019a}.
This builds on the success of the Advanced LIGO and Advanced Virgo first and second observing runs, which led to ten binary black hole detections \citep{AbEA2018b} and the detection of one binary neutron star (BNS) merger GW170817~\citep{AbEA2017b}.
The BNS detection was unique in many ways, including the observation of the electromagnetic signature of the ejected matter. This includes: 1.) Isotropic emission in the visible and near infrared of the dynamical ejecta following the coalescence called a ``kilonova'' (KN) counterpart, AT2017gfo \citep{ChBe2017,CoBe2017,2017Sci...358.1556C,2017Sci...358.1570D,2017Sci...358.1565E,2017ApJ...848L..25H,2017Sci...358.1579H,KaNa2017,KiFo2017,2017ApJ...848L..20M,2017ApJ...848L..32M,NiBe2017,2017Sci...358.1574S,2017Natur.551...67P,SmCh2017,2017PASJ...69..101U} 2.) the beaming emission of the relativistic ejecta with the short gamma-ray burst (SGRB), GRB170817A~\citep{AbEA2017e,GoVe2017} and 3.) the multi-wavelength afterglow due to interaction of the jet with the interstellar environment \citep{2017ApJ...848L..21A,LyLa2019,2018Natur.561..355M,2017Natur.551...71T}.
This event yielded a variety of results, including a measurement of the expansion rate of the universe \citep{2017Natur.551...85A,HoNa2018,CoDi2019}, limits on the equation of state (EOS) of neutron stars \citep{BaBa2013,AbEA2017b,RaPe2018,BaJu2017,CoDi2018,CoDi2018b}, and the likely formation of heavy elements \citep{JuBa2015,WuFe2016,KiFo2017,RoLi2017,AbEA2017f,RoFe2017,KaKa2019}.

The detection of the optical counterpart AT2017gfo \citep{2017Sci...358.1556C} at a distance of 40\,Mpc was helped by a three-detector gravitational-wave detection, constraining the final localization to $\approx$\,16 square degrees on the sky \citep{AbEA1019}.
With a sky localization of this size, there are a limited number of galaxies within the sensitivity volume of the gravitational-wave detectors, enabling a straight-forward search method of observing each galaxy for new objects. Similarly, synoptic survey strategies by larger field of view telescopes (e.g FOV $>$ 1 deg$^2$) was made easier by a localization of this size.
While this was remarkably good luck for the astronomical community, by far most of the gravitational-wave events before and since were one and two-detector observations, yielding much larger localization regions. Furthermore, with the two LIGO detectors having more than twice the sensitivity of Virgo during O3, and with different antenna pattern distributions over the three detectors, the two-detector observations will be the most likely case for any binary neutron star merger candidate in O3. This will effect the 1-50 BNS detections expected during O3, and will continue to be important into O4, when the number of expected detections varies between 4-80 per year \citep{AbEA2018}. Note that the angle-averaged binary neutron star range is already at 140 Mpc for LIGO Livingston (and about 120\,Mpc for LIGO Hanford), whereas available catalogs such as GLADE \citep{2018MNRAS.479.2374D} are only complete below $\sim$\,100 Mpc (although nearly complete at $\sim$\,150 Mpc).
Galaxy targeted follow-ups are significantly more limited in the case of binary black hole signals, which have generated interest for both potential gamma-ray (see e.g. \citealt{CoBu2016,VeDa2019}) and optical (see e.g. \citealt{SmCh2017}) counterpart searches.
For example, the first BNS detection candidate of the O3, LIGO/Virgo S190425z, was a single detector event with an initial sky localization from BAYESTAR \citep{SiPr2016} spanning $\sim$\,10,000 deg$^2$ at 155 $\pm$ 45 Mpc \citep{SiEA2019a} and an updated LALInference \citep{VeRa2015} skymap which reduced the localization region to $\sim$\,7500 deg$^2$ \citep{SiEA2019b}.
There were more than 50,000 galaxies inside in the 90\% volume for this source \citep{gcn24232}.

These large localizations have motivated many synoptic survey systems to search for optical counterparts. These include the Zwicky Transient Facility (ZTF) \citep{Bellm2018,Graham2018,DeSm2018,MaLa2018}, Palomar Gattini-IR (\citealt{Moore2019}, De et al. in prep.), the Dark Energy Camera (DECam) \citep{FlDi2015}, the Gravitational-wave Optical Transient Observer (GOTO) \citep{Ob2018}, the Panoramic Survey Telescope and Rapid Response System (Pan-STARRS) \citep{KaBu2010}, the All-Sky Automated Survey for Supernovae (ASASSN) \citep{ShPr2014} the Asteroid Terrestrial-impact Last Alert System (ATLAS) \citep{ToDe2018}, the Rapid Action Telescope for Transient Objects (TAROT) \citep{2008PASP..120.1298K} and the MASTER global robotic network \citep{2010AdAst2010E..30L} amongst many others.

While the use of synoptic systems was already typical during the first and second observing runs, there is a growing trend of telescope ``networks,'' some of which are built around these synoptic systems. These networks use various facilities to perform rapid follow-up and classification of objects (see Figure~\ref{fig:flowchart}). For example, ZTF, Palomar Gattini-IR, DECam, and the GROWTH-India telescope\footnote{https://sites.google.com/view/growthindia/} (Bhalerao et al., in prep.) are scheduled by the Global Relay of Observatories Watching Transients Happen (GROWTH\footnote{http://growth.caltech.edu/}) network \citep{CoAh2019b} (in addition to predominantly galaxy-targeted follow-up systems such as the Kitt Peak EMCCD Demonstrator (KPED) on the Kitt Peak 84 inch telescope, \citealt{Coughlin2018}). In addition, the Global Rapid Advanced Network Devoted to the Multi-messenger Addicts (GRANDMA) uses small to medium sized telescopes spread over the entire globe, comprised of over 20 classical and robotic facilities \citep{GRANDMApaper}.
These networks are useful for a few reasons. Due to the considerable size of the sky localizations, it is advantageous for each search to utilize telescopes capable of covering both hemispheres. In addition, coordinated observations can save precious target of opportunity time on large aperture systems. Once the region has been imaged and candidates are identified, having worldwide coverage allows for continuous follow-up of candidates. This coverage enables identification and characterization of potential counterparts at high cadence and with multi-band photometry. Having detections as early as possible is important for understanding the source mechanisms \citep{Arc2018}.

There has been much recent interest in optimizing the methods to schedule observations, given the use of significant telescope time to search and follow-up electromagnetic counterparts. 
The gravitational-wave counterpart search effort, with the gamma-ray burst and neutrino counterpart searches closely related \citep{SiCe2013,CoAh2019}, is unique in the community, given the significant search regions requiring coverage. Recently, an open-source codebase named \texttt{gwemopt}\footnote{https://github.com/mcoughlin/gwemopt} (Gravitational Wave - ElectroMagnetic OPTimization, \citealt{CoTo2018}) was developed, deriving concepts from the community on how to optimize optical follow-up of gravitational-wave skymaps. This includes information about how the telescopes should tile the sky, allocate available telescope time to the chosen tiles, and schedule the telescope time. We have developed generic algorithms to handle these tasks that would be useful for a wide variety of telescope setups; this includes telescope placement on the Earth, as well as their instrument configurations, including field of view, filters, typical exposure times, and limiting magnitudes.

While single telescope optimization remains important, it is clear that methods extending some of these methods are required for network level optimization. 
In this paper, we will introduce two basic extensions to the single telescope model of follow-up, which we call ``iterative'' and ``overlapping.'' The idea is to make straightforward extensions to the single telescope scheduling models, which have generally been shown to be robust and successful \citep{CoAh2019,CoAh2019b,AnGo2019}, including during the 190425z follow-up performed by GROWTH and GRANDMA \citep{gcn24227,gcn24187,gcn24191,CoAh2019b}. We note that the techniques are generic enough to be used with different scheduling algorithms, some of which we will describe below.

\section{The Iterative Algorithm}

We will briefly review the state of the art in single-telescope scheduling most relevant for a multi-telescope network. One can broadly break up the process of scheduling into three categories: 1.) How the telescopes should tile the sky, 2.) How the telescope array should allocate time to each tile, and 3.) How to schedule that time between telescopes. Ideally, all three of these would be done at once, as of course, the ability to schedule a tile should inform how much time is possible to allocate to it. In practice, \texttt{gwemopt} simply removes any tiles that are not observable during the time requested,``good enough'' to create sensible schedules, but still suboptimal relative to a schedule that optimizes all three simultaneously.

Despite its simplicity, this approach allows for the creation of tiles the size of the telescope's field-of-view with minimal overlap covering the whole sky. 
This is typically done using the ``hierarchical'' and ``greedy'' methods \citep{GhCh2017,CoTo2018}; the idea of both of these tiling schemes is to decrement to zero the probability in the map enclosed in any already placed tile. In other words, each tile placed leads to a change in the skymap that is being tiled, where the locations in the map covered by that tile is set to zero. This prevents, for example, the possibility of double counting the probability contributed by a particular sky location when multiple tiles cover the same location.
For some instruments, such as ZTF, the tiles are pre-determined to simplify difference imaging. ZTF in particular has both a ``primary'' and ``secondary'' grid, where the two grids are designed to fill in the $\sim$\,15\% of the field of view that is not imaged due to gaps in between the individual CCDs. Part of the ``job'' of the scheduling software is to optimize the use (or not) of overlapping tiles like those of ZTF.
In general, taking images in the secondary grid has not been a priority for ZTF, and therefore there are not references for all of the fields in this grid; for this reason, it will be useful to have methods to fill-in these regions with other systems.

The most important metric for any tile is the integrated spatial probability of a gravitational-wave source lying within it.
This is computed by using the gravitational-wave skymaps, which report either the 2D probability $L_\textrm{GW}(\alpha,\delta)$, in right ascension $\alpha$ and declination $\delta$, or 3D probability, which includes probability distributions for the luminosity distance $D$ as a function of sky location (for explanation see the LIGO-Virgo user guide\footnote{https://emfollow.docs.ligo.org/userguide/}). The integrated probability in a tile is computed as a double integral over right ascension and declination
\begin{equation}\label{eqn:tile_ranked}
T_{ij} = \int_{\alpha_i}^{\alpha_i+\Delta \alpha}\int_{\delta_i}^{\delta_i+\Delta \delta}L_\textrm{GW}(\alpha,\delta)d\Omega.
\end{equation}
In general, a fiducial target integrated probability, usually around $90$\%, is used to determine the number of tiles to consider for imaging and scheduling.

\begin{figure}[t]
 \includegraphics[width=3.3in]{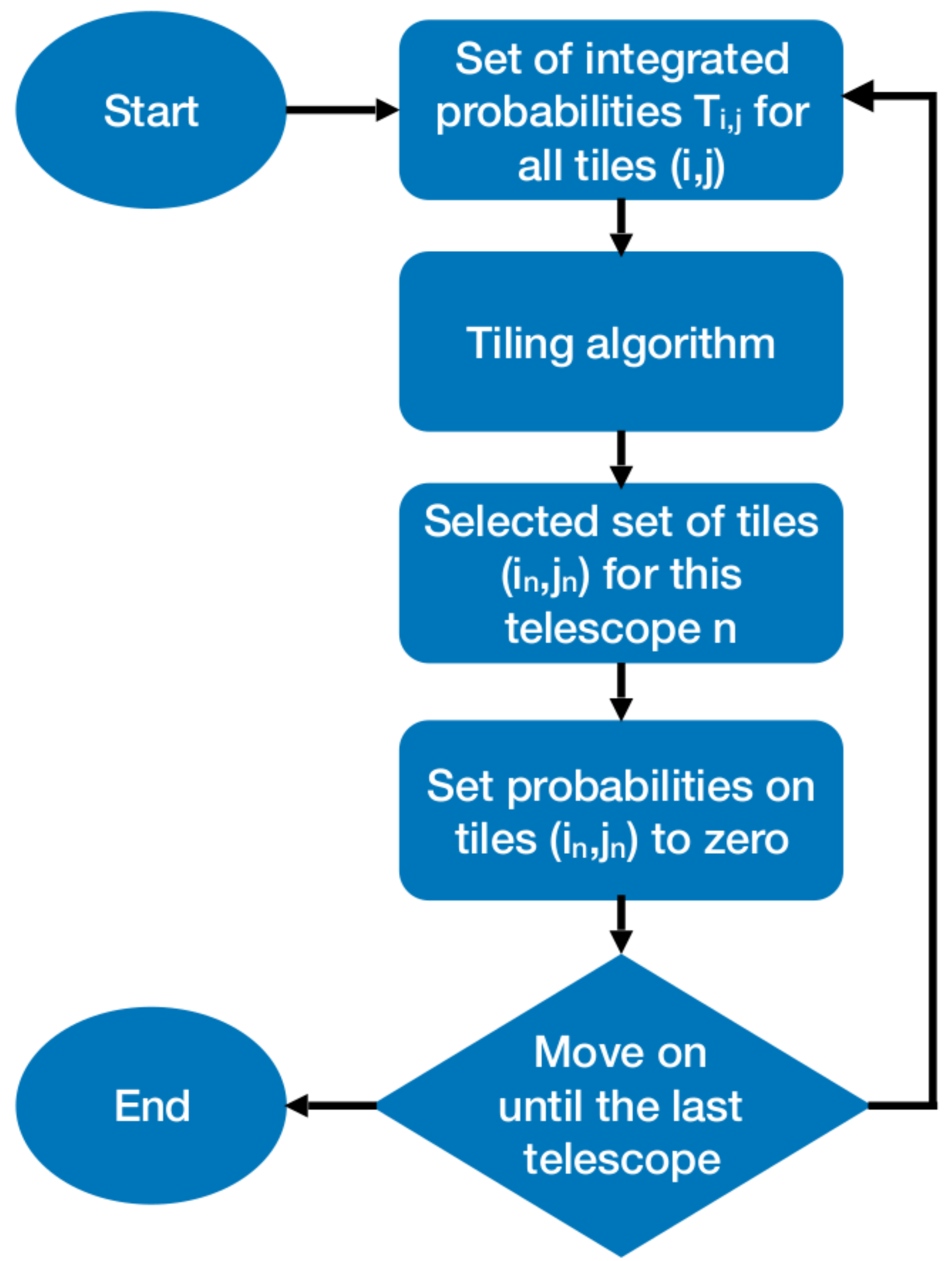}
  \caption{Flowchart of the ``iterative'' algorithm presented in the text.}
 \label{fig:iterative_flowchart}
\end{figure} 

This brings us to our first innovation in the tiling method, where we will use the decrementing scheme from the ``hierarchical'' and ``greedy'' methods to expand to a multi-telescope network\footnote{``hierarchical'' determines the locations of the tiles one at a time, while ``greedy'' optimizes all locations simultaneously}.
The first telescope in the network is scheduled as usual, yielding an optimal set of tiles for that telescope.
After the first telescope is scheduled, the gravitational-wave skymap is decremented with all of the pixels covered by the first telescope's observations set to zero.
Following that, the tiling for the second telescope is computed with this modified map, and the process continues.
At the end, this yields a map covered by tiles in the telescope network with minimal overlap.
This algorithm is summarized in Figure~\ref{fig:iterative_flowchart}.

The first telescope scheduled should likely be the ``best'' telescope in the network, i.e. the most sensitive and/or most reliable transient finder.
Another option for the first telescope is the one that simply has the most localization probability observable by that site at the time of the trigger, and this option has been added as a flag to \texttt{gwemopt}.
Different operators may, of course, determine what ``best'' means for them, but it is important as this telescope is likely, depending on placement on the Earth, going to be tiling the regions of highest likelihood in the skymap.

There are many considerations for what constitutes best here. Objectively, ``etendue'' is a reasonable metric, which is the product of the aperture area of the primary mirror times the field of view covered, and so the units are degrees squared times meters squared. The Large Synoptic Survey Telescope \citep{Ivezic2014} design is famous for optimizing around this quantity. But experience has shown that observatories whose observations strongly constrain how recently an object appeared, such as from ZTF, ATLAS, and Pan-STARRS, are incredibly important for limiting the number of objects that require follow-up. For example, while ZTF has been reporting $\sim$\,20 objects per event, DECam follow-up has yielded an order of magnitude or more due to its lack of recent limits \citep{AnGo2019,GoAn2019}. For this reason, it is not necessarily obvious that etendue is the deciding proxy for ``best.'' For example, the telescope ordering can vary depending on a variety of metrics: 1.) the significance of the event, 2.) the nature of the alert, 3.) the size of the gravitational-wave sky localization, 4.) the distribution of the telescopes around the globe that will be used for follow-up, 5.) the available filters in a given system, and 6.) the delay between the trigger time and the start of observations.

\begin{figure*}[t]
 \includegraphics[width=3.5in,height=3.0in]{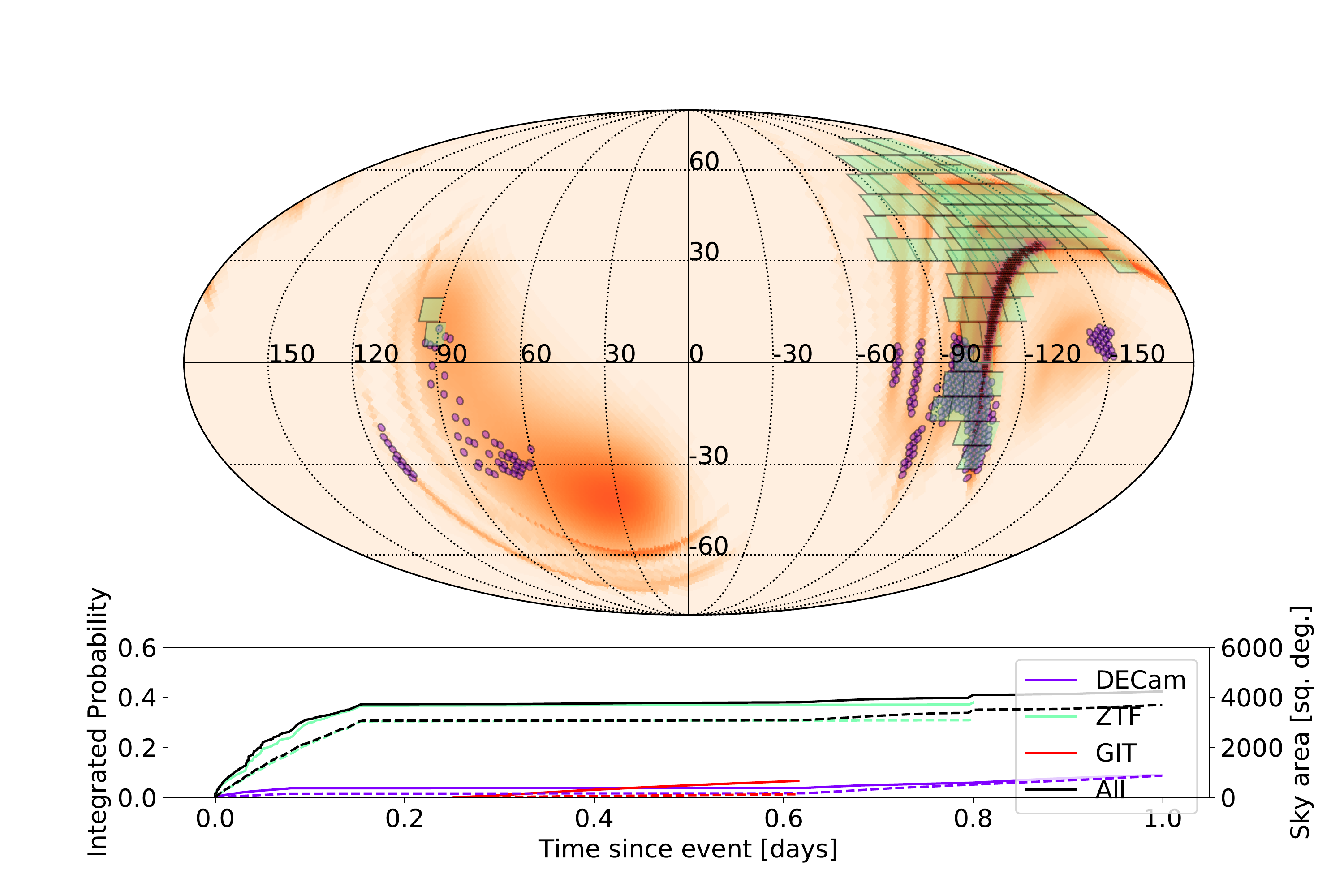}
 \includegraphics[width=3.5in,height=3.0in]{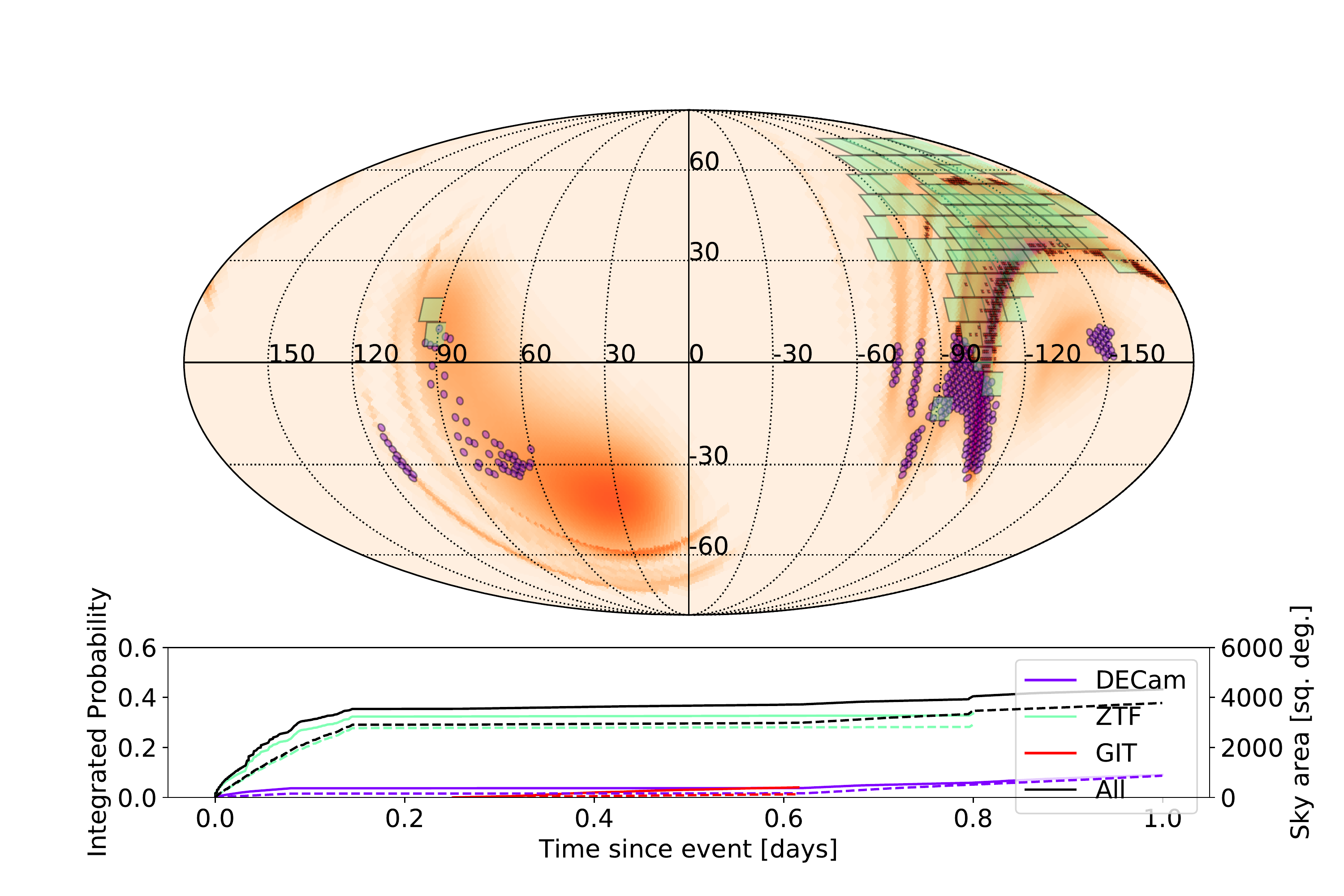} 
  \includegraphics[width=3.5in,height=3.0in]{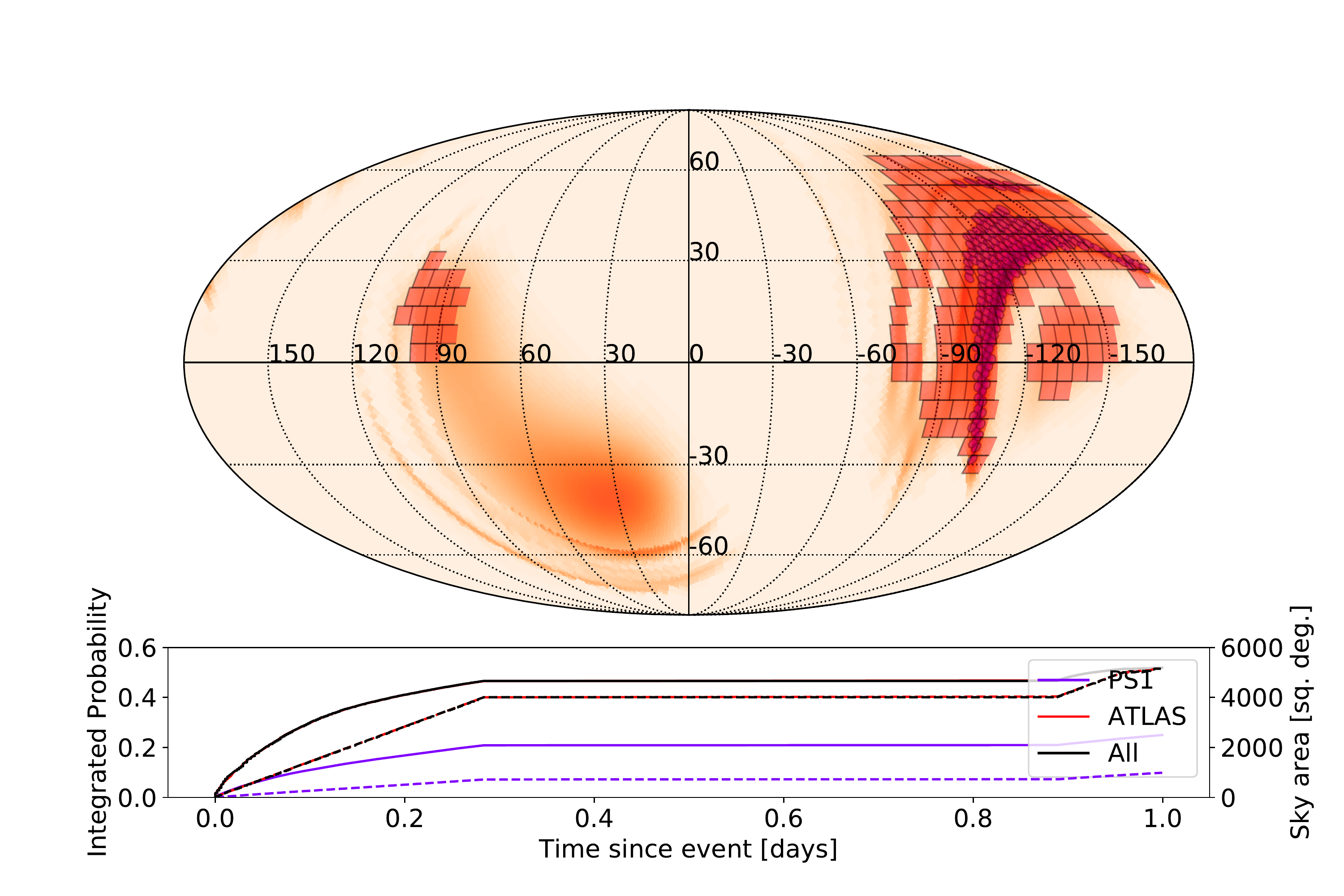}
 \includegraphics[width=3.5in,height=3.0in]{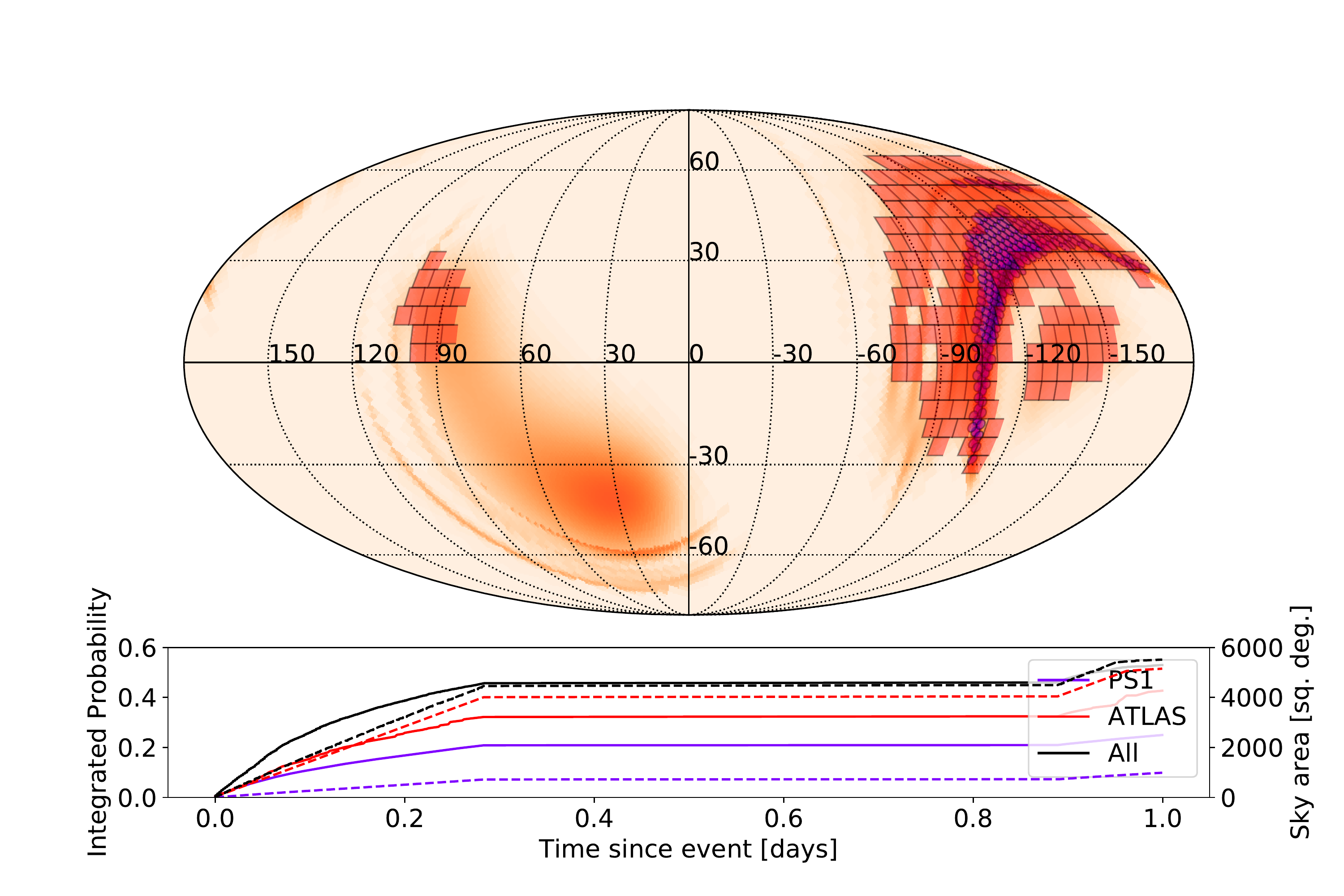} 
   \includegraphics[width=3.5in,height=3.0in]{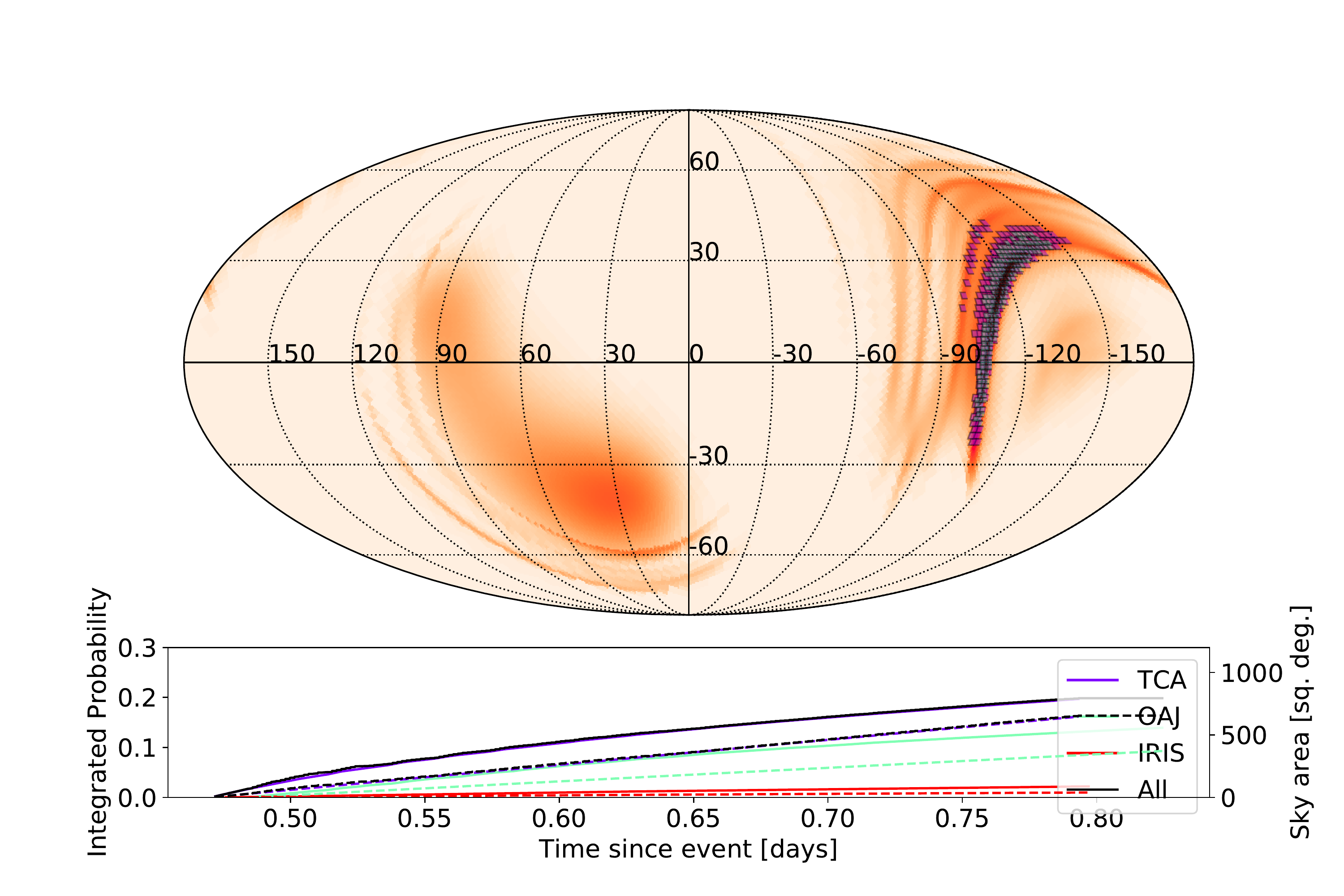}
 \includegraphics[width=3.5in,height=3.0in]{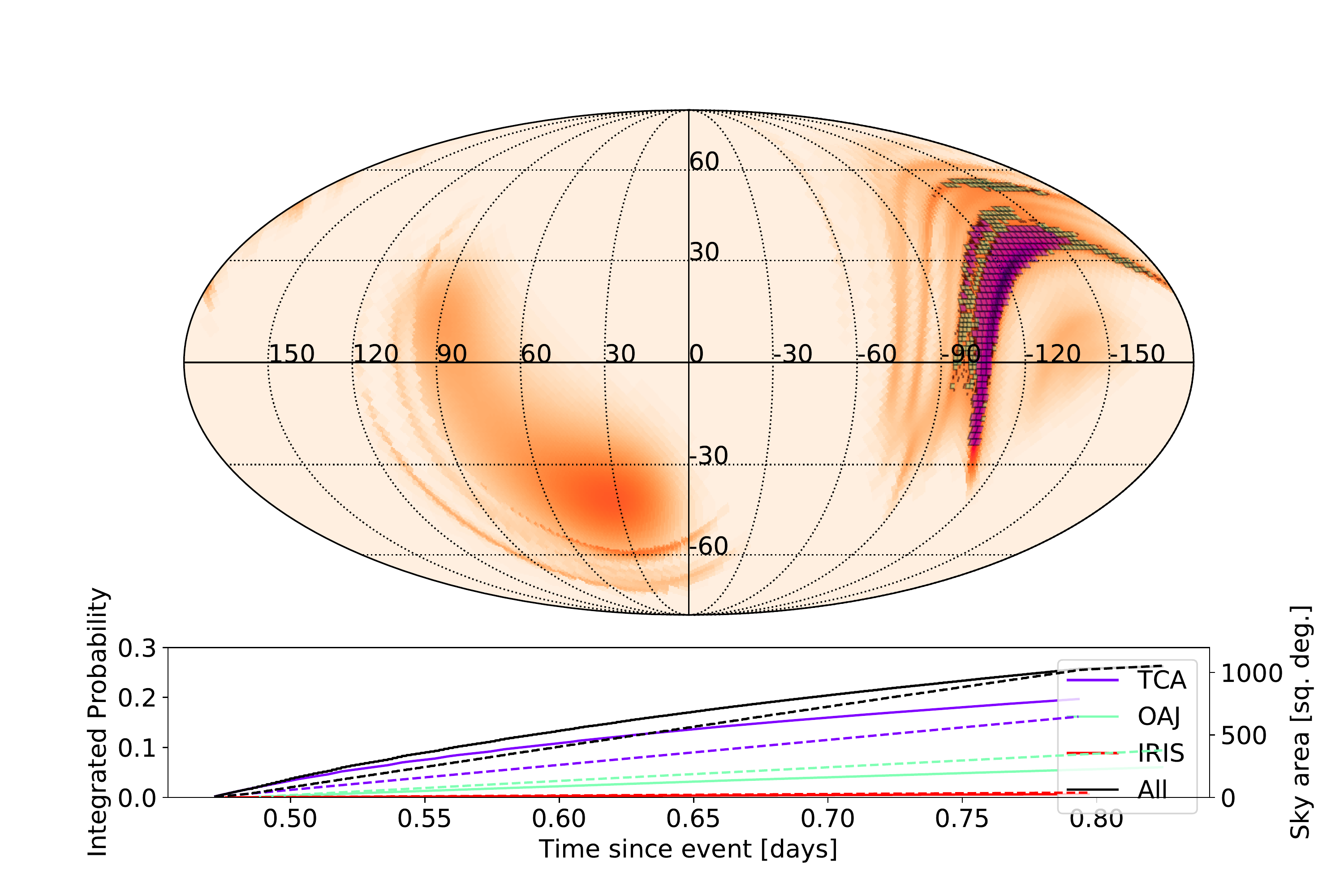} 
  \caption{Optimized coverage of S190425z. The top row shows the optimization for the ``GROWTH'' network, which includes ZTF, DECam, and GROWTH-India tiles. The left shows the tiles drawn using the original scheduling algorithm, while the right is the same for the iterative method discussed in the text. The middle row shows the same for the Pan-STARRS and ATLAS pair. The bottom row shows the same for some telescopes of the ``GRANDMA'' network, with IRIS, OAJ, and TAROT-Calern. The dashed lines indicate the sky area covered, while the solid line indicates the integrated probability covered of the 2D skymap.}
 \label{fig:iterative}
\end{figure*} 

\begin{figure*}[t]
 \includegraphics[width=7.0in]{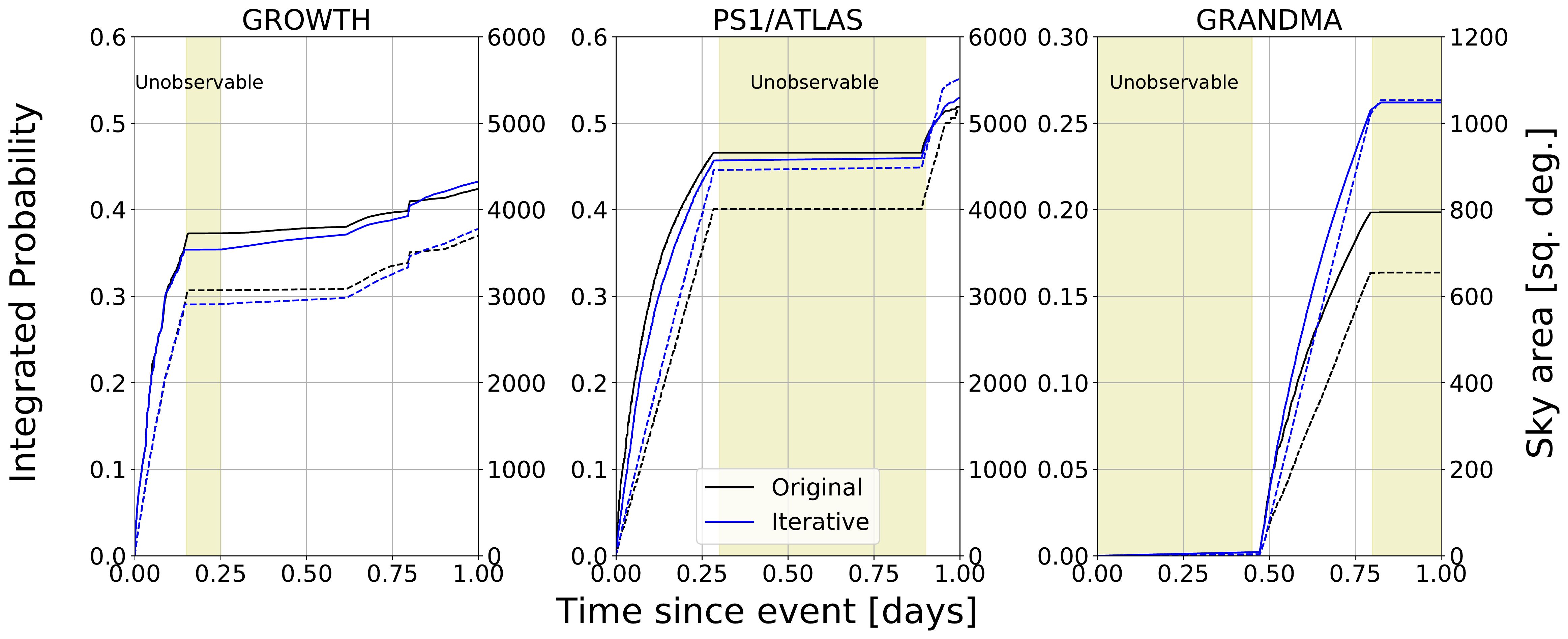}
  \caption{Optimization comparison for the GROWTH, Pan-STARRS and ATLAS, and GRANDMA networks. The dashed lines indicate the sky area covered, while the solid line indicates the integrated probability covered of the 2D skymap. The black lines correspond to the original scheduling algorithm while the blue lines correspond to the iterative method. The yellow band is the approximate ``daytime'' for the network, where no observations are taken for this particular skymap.}
 \label{fig:iterative_multi}
\end{figure*}

In order to compare the original and iterative methods, we highlight a few examples of existing telescope networks.
First of all, we include a sample ``GROWTH'' network \citep{CoAh2019b}, which includes ZTF, DECam, and GROWTH-India tiles.
We also include the Pan-STARRS and ATLAS pair, currently scheduled and analyzed by the same teams out of the University of Hawaii and Queen's University Belfast.
Finally, we include a sample of the ``GRANDMA'' network \citep{GRANDMApaper}, which includes the IRIS 50\,cm telescope\footnote{http://iris.lam.fr/}, the Observatorio Astrofísico de Javalambre (OAJ) 80\,cm\footnote{https://oajweb.cefca.es/telescopes/jast-t80}, and TAROT-TCA located at Calern Observatory\footnote{http://tarot.obs-hp.fr/}.
The diversity of the tested networks gives a first indication of the benefits of iteration; we employ 1.) a pair of telescopes with similar $\sim$\,1-deg$^2$ field of views (FOV) in the same region with OAJ and TAROT, 2.) a pair of telescopes with very different FOV with Pan-STARRs and ATLAS, located at the same observatory, and 3.) a pair of telescopes with ZTF and DECam, with different FOV located at different latitudes.

We show a comparison between the original and iterative methods with S190425z sky localization area in Figure~\ref{fig:iterative}.
While S190425z is taken as a single example, the results can change qualitatively between, for example, sky maps with large and small sky areas, although this method is appropriate for both.
To guide the reader's eye for interpreting the plots in Figure~\ref{fig:iterative}, we note that the cumulative area and probability covered are the same in both the left (original) and the right (iterative) panels.
The black lines in each plot display the integrated probability based on the sum of the contributions of the telescopes in the network.
Significant differences between the black lines and the lines from the individual telescopes indicate that the individual telescopes are imaging different areas of the sky (i.e. accumulating probability and sky coverage complementary to one another).
On the left hand side, there is usually one telescope (with the largest field-of-view) that tracks most closely to the overall line, while on the right hand side, there is clear separation.
The improvement is most clear at early times, when telescopes can image different portions of the sky localization that still have significant probability.
As expected, the iterative method covers both more S190425z sky localization and larger cumulative probability than independent scheduling of the individual telescopes. 
We show a direct comparison of the cumulative area and probability in Figure~\ref{fig:iterative_multi}.

For the GRANDMA network, the skymap coverage of S190425z is nearly doubled using the `iterative'' tiling from 660 to 1060 square degrees, with a total cumulative probability improving from $\sim$\,0.20 to $\sim$\,0.26.
The effect is also visible for the Pan-STARRS and ATLAS pair, even with the different field of views. The ``iterative'' tiling increased the total coverage by $\sim$\,1000 square degrees. 
Indeed, the integrated probability and cumulative sky area covered is $\sim$\,0.51 and $\sim$\,4840 square degrees with the original scheduling, while the integrated probability and cumulative sky area covered is $\sim$\,0.55 and $\sim$\,5735 square degrees with the iterative scheduling (see Figure~\ref{fig:iterative}). Note that identifying the overlapped tiles can also support multi-band observation of the kilonova. Finally, the DECam-ZTF pair is naturally distributed between the north and south. 

However, in the case of S190425z, the highest probability pixels were located in the sky observable by both sites. The method helps again to distribute the skymap observation more efficiency than the original tiling case. The margin of improvement of the method presented here depends strongly on the gravitational-wave localization area, but already in these three cases we see improvement. In addition, the contributions of the smallest field of view telescopes in each network is increased since the iterative method assigns yet unexplored sky coverage to any telescope in the system.

\begin{figure*}[t]
 \includegraphics[width=3.5in,height=3.0in]{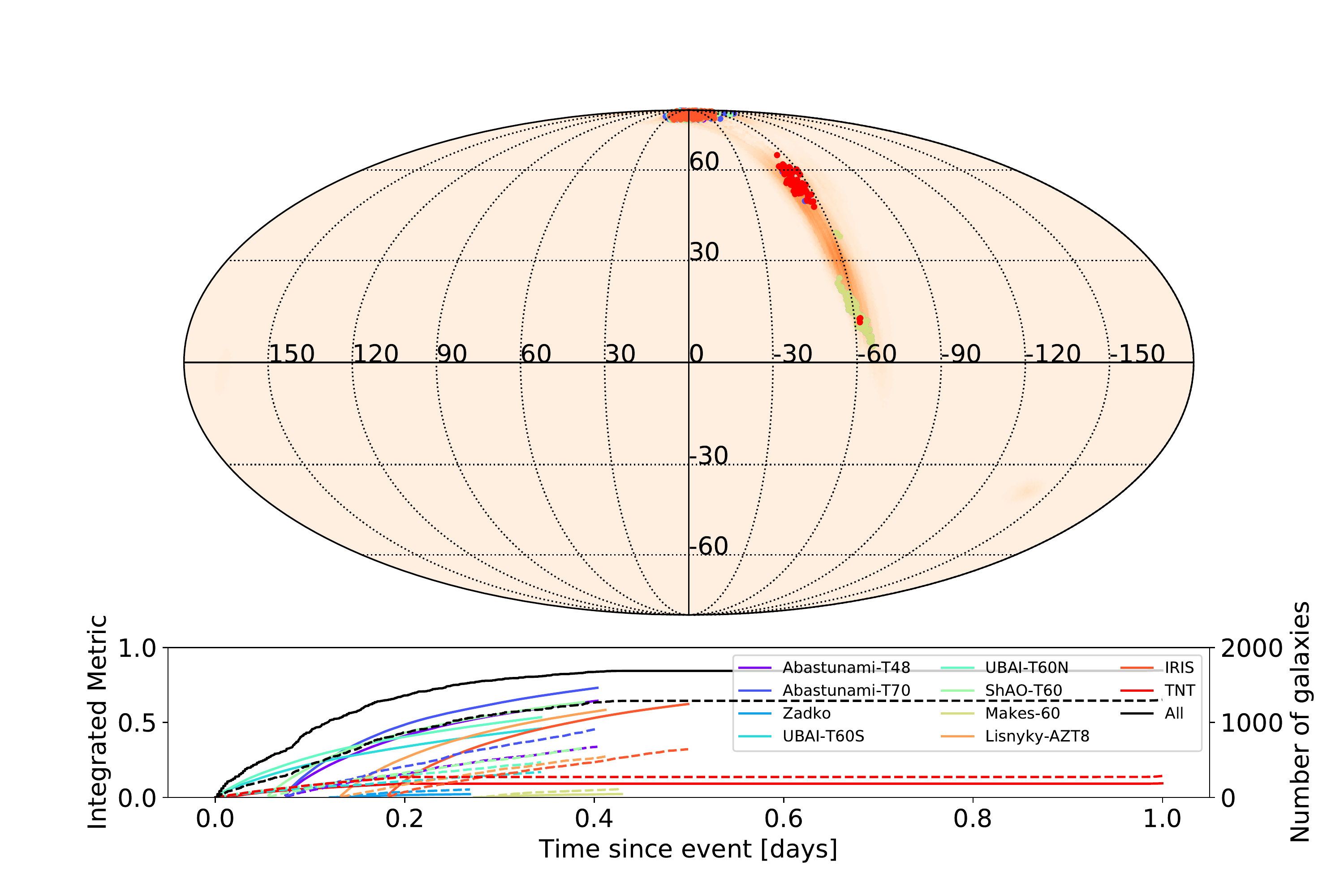}
 \includegraphics[width=3.5in,height=3.0in]{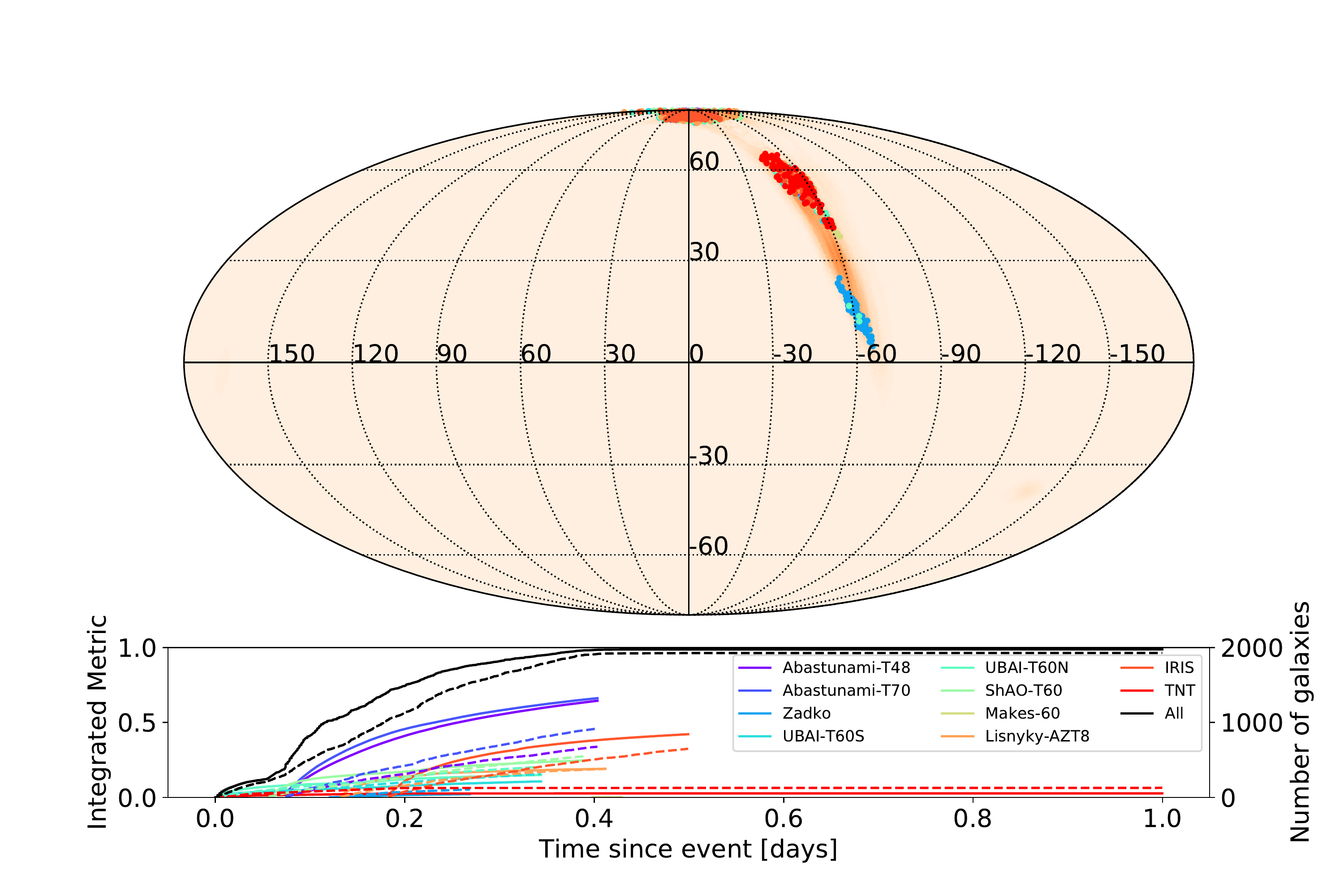}
  \caption{Optimized coverage of S190426c using the GRANDMA ``galaxy targeting'' Network. On the left is the original algorithm where the telescopes are scheduled separately, and on the right, where they are scheduled iteratively. The total cumulative metric covered improves from 85\% to 99\%, and the total number of galaxies imaged improves from 1303 to 1929.}
 \label{fig:galaxy}
\end{figure*} 

The extension of this method to the galaxy-targeting approach is straight forward. The idea of the galaxy-targeted method is to use catalogs such as GLADE \citep{2018MNRAS.479.2374D} or the Census of the Local Universe (CLU) catalog \citep{CoKa2017} that list the galaxies within the gravitational-wave localization volume; the reduction in area requiring coverage makes it possible to use small field of view instruments in the counterpart searches effectively.
A variety of metrics exist for determining which galaxies should be imaged (see \cite{ArMc2017} for an example).
They usually include a proxy for the location of the galaxy within the gravitational-wave localization volume and the galaxy's mass or star formation rate. They sometimes also account for the specific telescope's sensitivity to a given transient at the galaxy's distance to avoid pointing at objects likely to be too far away to be detectable.
In any case, for galaxy targeting, the idea is to simply remove the galaxies that have otherwise been scheduled by previous telescopes. 
Similar to the synoptic case, after the first telescope is scheduled, the weight associated with a given galaxy that is scheduled is set to zero.
From then on, future telescopes in the network will no longer attempt to schedule that galaxy given its weight, and instead will schedule others. As an example, we use use this method to schedule eleven telescopes of the GRANDMA network, each with a field of view $<$ 1 deg$^2$, on the S190426c sky localization, a neutron star-black hole candidate with a 90\% credible region covering 1260 deg$^2$ at a distance of 375 $\pm$ 108 Mpc \citep{ChEA2019a}. Figure~\ref{fig:galaxy} shows that the improvement in both the number of galaxies and the metric. More concretely, the total number of galaxies imaged improves from 1303 to 1929. For the galaxy targeted portion, the total cumulative adopted metric covered improves from 85\% to 99\%, where we have only included galaxies inside of the 90\% contour when computing this number.

One of the major downsides to this schema overall is the possibility of weather related failures, and the loss in opportunity of imaging high likelihood regions of the sky. This leads to the idea of ``golden tiles,'' which are regions of the sky that are not decremented at each step. Because the inner portion of the sky localization region (say the inner 50\%) tends to be much smaller than the outer portion of the sky localization region (say the outer 90\%) \citep{SiPr2014}, the opportunity cost of imaging the higher likelihood region is lower than the outer region. For this reason, we implemented a user-definable inner percentage of the skymap that is not incremented at each step. Based on the dimension of the telescope network, and the distribution of the telescopes over the globe, one idea would be to create coverage in groups of telescopes at similar longitudes. In each group, the optimization of the scheduling for systems at different latitudes would be performed with the method presented here. For example, if scheduled in four groups, it would allow for a maximum of a six hour gap of observation of the same target in the sky. It seems most appropriate to have some redundancy with the golden tiles, but with different filters, to create a balance between the possibility of kilonova detection, which requires maximum coverage, and time evolution of the kilonova in multiple filters. The percentage of the golden tiles will then be a combination of time allocation for each telescope to follow-up gravitational alerts and the size of the sky localization area. In this sense, the threshold can be based on percentage of the sky that can be covered.

Once the preferred set of tiles has been scheduled, how to allocate exposure times is the next question for the scheduling algorithm to address. We will not discuss this issue much herein, as in general this decision is very telescope and observer dependent, and has been discussed extensively elsewhere (see \cite{CoTo2018} for a summary).
In general, one popular proposed technique is to allocate exposure times proportional to the probability enclosed in a tile (see, for example, \cite{CoSt2016a} or \cite{GhCh2017}).
It is also possible, with the information in three-dimensional sky localization, to image deep enough in each field to reach a fixed absolute magnitude. In general, this would mean taking longer exposures in the center of the localization and shorter exposures at the edge of the skymap. In practice, this makes it much more complicated to compare objects found in the search. The pre-merger limits, especially those derived from a survey that uses a fixed exposure time, will generally be similar across the localization. Changing exposure times within a search will lead to significant differences in how objects are determined for follow-up.
Given the desired redundancy described previously, it may also be appropriate to choose a uniform maximum magnitude based on the size and distance estimate of the localization, and then adjust the exposure time for each telescope to achieve that specific depth. For example, for a given localization, if ZTF could use 90\,s exposures to cover the desired area, achieving a depth of $\sim$\,21, then GROWTH-India might also adjust its exposure time to achieve that same depth.

Once time has been allocated to each telescope's tile, the next task is to schedule the observations. In most cases, there is insufficient time during the night to perform all observations, and therefore the job of the scheduling software is to optimize the subset actually taken.
In general, the optimization schemes employed in \texttt{gwemopt} \citep{CoTo2018} weight each tile by a combination of its integrated probability (or metric in case of the galaxy targeting) and its current and future visibility, so as to schedule as many tiles as possible while maximizing the probability covered. As part of this, the scheduling software must also account for practical constraints such as altitude and moon-sky brightness limits.
There are also more telescope-specific issues such as slew rate, hour angle constraints, and camera readout times. 
One of the most important aspects to the scheduling is the cadence at which a telescope returns to a particular field or galaxy, and the filters that are chosen when imaging.
For instance, ZTF has mainly been imaging in $g$-$r$-$g$ band exposure blocks to ensure a measurement of both color and potential change in luminosity in a single band \citep{CoAh2019}.
The observations are separated in time, not only to measure a luminosity change, but also in order to reject asteroids and other transient objects.
The scheduling algorithms used to address these issues are examined thoroughly in \cite{CoTo2018}, and we refer the reader there for further information

\section{The Overlapping Algorithm}

While it is convenient for each telescope to be its own follow-up resource, given the sky localizations involved and the likelihood of having tiles set at certain longitudes, it is useful to potentially image a location on the sky with multiple systems. The problem is that independent scheduling will lead to the highest probability region being imaged around the same time (at least in the case where the sites are at similar locations). In other cases, especially when the localization area is very large or the number of telescopes in the network small, it makes sense to temporally separate observations of the same field by different telescopes.
For example, a patch imaged by DECam might be visible 6 hours later with ZTF, so it is not necessary to to wait until the second night of DECam observations to get a second epoch if ZTF also images that location.

\begin{figure}[t]
 \includegraphics[width=3.5in]{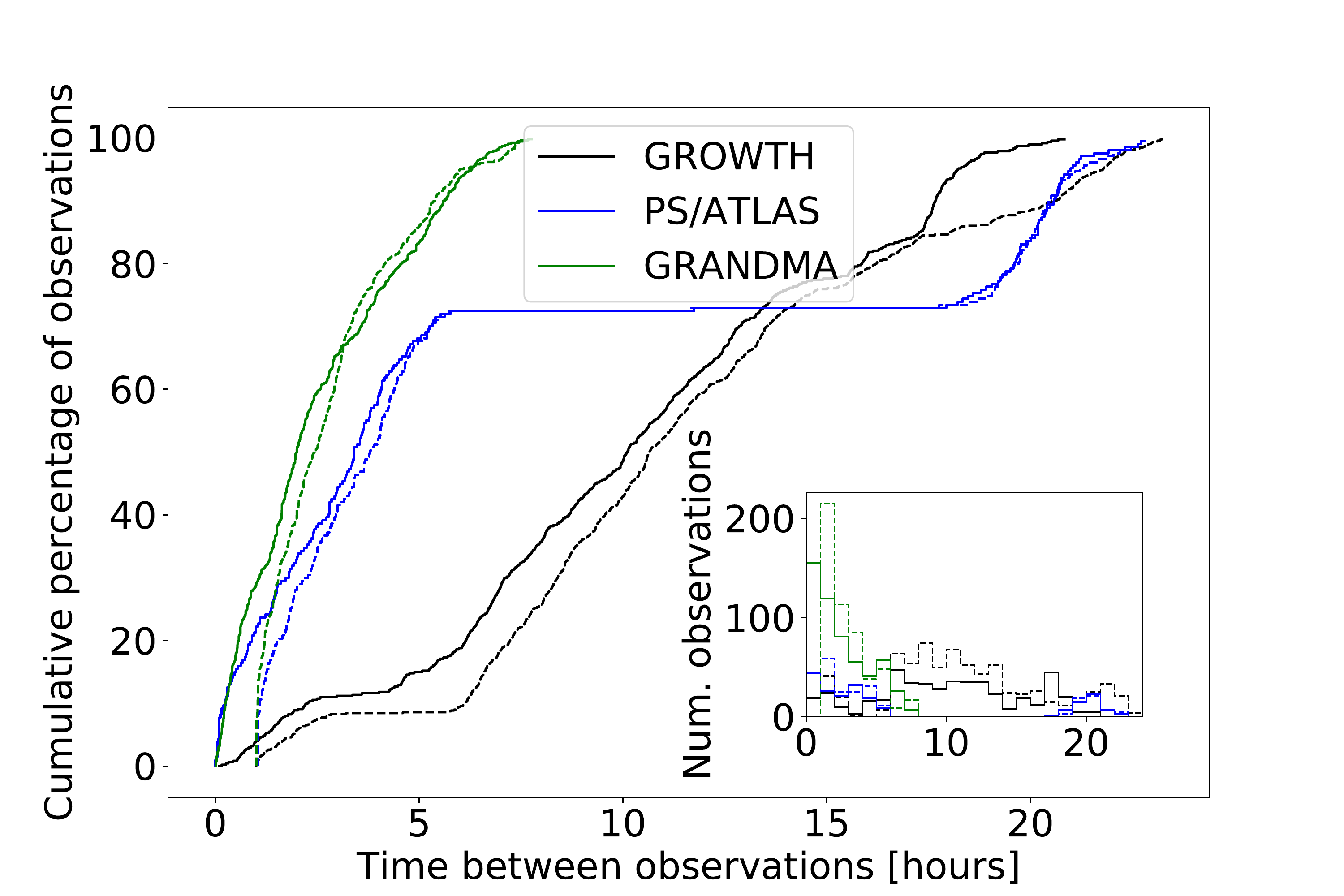}
  \caption{Cumulative histogram of the difference between telescope observations of the same patch of the sky for S190425z. We plot the original algorithm in solid and overlapping algorithm in dashed. The inset shows the original histogram. The reader should note the lack of observations within the same hour in the ``overlapping'' traces, shown in dashed for the various networks.}
 \label{fig:overlapping}
\end{figure} 

To enable this, we once again modify the existing scheduling algorithms in a simple way.
Similar to the ``iterative'' method, we allow the first telescope to schedule its observations as before.
For the next telescope, we specify a minimum difference in time (or time delay) between observations in a given field.
Algorithmically, we impose this observability constraint on a given field, just like is done for moon proximity or airmass, such that it will not be scheduled within the specified time delay of a previous observation.
In the example that follows, we choose one hour, more than sufficient to differentiate between real transients and asteroids, as well as potentially measuring a change in luminosity.
This is appropriate, for example, for a kilonova, which is expected to show a rapid evolution in magnitude \citep{Me2017}; GW170817 faded $\Delta r \sim 1$\,mag per day over the first 3 days post explosion.
If the goal is to simply measure a change in luminosity over these time-scales, as opposed to discard asteroids as potential candidates, something like a 6\,hr delay may be more appropriate.

The difference between the normal scheduling and the ``overlapping'' scheduling is shown in Figure~\ref{fig:overlapping}.
We show both the histograms and the cumulative version.
We would like the reader to note the lack of observations within the first hour in the ``overlapping'' schedule case, as is expected.
We also note that the number of observations in overlapping fields rapidly ``catches up'' after that first hour, since the software will optimize around scheduling the highest probability fields once the one hour constraint has been lifted.
In this way, the algorithm is successful at not only preventing overlapping observations of the same field within the time-frame specified, significantly limiting the false positives that arise from asteroids, but also does not prevent that field from being observed by that system at a later time because, for example, that particular field had set. 

\section{Conclusion}

In this paper, we have described straightforward strategies for a multi-telescope network.
We have shown how the introduction of two new tiling and scheduling schemes make it possible to use existing single-telescope strategies in a network capacity.
This work builds upon examples of optimization in this regard done ``by hand'' previously \citep{gcn24316}.
Algorithms of this type open up the possibility of coordinated observations between telescopes for gravitational-wave follow-up.
This also brings to bear a variety of the scheduling techniques that have been developed.
For example, the slew-optimized scheduling algorithms that have recently developed \citep{RaAn2019} are likely more important in a network where the tiles scheduled are likely more spatially separated.
It also opens the possibility of prioritizing different schedules with different telescopes. For example, one telescope might use an airmass-based optimization to maximize the sensitivity given a fixed exposure time, at the cost of not imaging the highest probability tiles as early as it might otherwise. This might be more palatable when another system is using a basic greedy algorithm to image the highest probability tiles as early as possible.

Longer term, we wish to design optimizations that will vary the number of fields accounting for scheduling constraints, instead of separating the steps of choosing fields, allocating time to them, and then scheduling.
We also want to define the metrics generically enough that having more than one instrument is like having a more sensitive version of a single instrument, which is difficult, given the constraints that the telescopes are scattered around the Earth, with varying fields of view and sensitivities, etc.
It should also be the case that the choices a human would intuitively make, such as using different telescopes to cover disparate parts of the localization when they are significantly separated on the sky to save slew time, are naturally accounted for in these metrics.
It should be the case that the use of multiple telescopes correctly should be able to reduce slew time by minimizing the size of their patches.
In addition, not relying on an ordered list of telescopes should result in even better sky coverage.
There is also the open question of how a network should be optimally used, when accounting for the available time for target of opportunity observations.
For example, in our analysis, we have assumed that taking complete control of each of these systems for the night following the event is appropriate. In practice, the most desirable systems should not / cannot be used on all events, given that their time is limited.
Determining criteria for their optimal use should be a focus of future work.

As time goes on, the detection of kilonovae should become more frequent; this will include searching short gamma-ray burst counterparts for kilonova signatures \citep{AsCo2018,GuZi2019}.
Thus the future perspective will be to adopt scheduling designed for studying the physical mechanisms at stake, not only for detecting the kilonova (see e.g. \citealt{AnAn2019}).
Then, instead of focusing on such things as ensuring we image an object twice in a given amount of time, we can instead prioritize metrics that are kilonova science targeted.
This is where the  ``overlapping'' scheme may become important, as using multiple telescopes can turn a simple detection into a discovery by measuring rapid changes in color and/or luminosity that may be difficult on a single system.
For example, it might be useful to do a second or third round of imaging instead of exploring more of the probability volume; this may lower the odds of detecting a kilonova, but would increase the science output if it is present.
The set of filters to use can be optimized depending on whether the adopted strategy prioritizes color or luminosity variations.
In addition, there are science cases for the detection as early as possible, and even for ``non-detections'' in the early photometry \citep{Arc2018}.

In conclusion, in the open source software \texttt{gwemopt} \citep{CoTo2018}, we have implemented a first optimization of a network level follow-up of gravitational-wave alerts, showing the substantial gains that can potentially be made by coordinated scheduling of existing telescope networks.

\acknowledgments

M.~W.~Coughlin is supported by the David and Ellen Lee Postdoctoral Fellowship at the California Institute of Technology. Sarah Antier is supported by the CNES Postdoctoral Fellowship at Laboratoire Astroparticle et Cosmologie. David Corre is supported by a CNRS Postdoctoral Fellowship at Laboratoire de l'Accélérateur Linéaire.
N. Christensen and T. Mikulski acknowledge support from the National Science Foundation with grant number PHY-1806990, and T. Mikulski also acknowledges support from the Towsley fund at Carleton College.
The UCSC (D. Coulter, R. Foley, D. Reed) team is supported in part by NASA grant NNG17PX03C, NSF grants AST-1518052 and AST-1911206, the Gordon \& Betty Moore Foundation, the Heising-Simons Foundation, and by a fellowship from the David and Lucile Packard Foundation to R.J.F.

\bibliographystyle{aasjournal}
\bibliography{references}

\begin{thebibliography}{}
\expandafter\ifx\csname natexlab\endcsname\relax\def\natexlab#1{#1}\fi
\providecommand{\url}[1]{\href{#1}{#1}}

\bibitem[{{Aasi et al}(2015)}]{aLIGO}
{Aasi et al}. 2015, Classical and Quantum Gravity, 32, 074001

\bibitem[{{Abbott} {et~al.}(2017){Abbott}, {Abbott}, {Abbott}, {Acernese},
  {Ackley}, {Adams}, {Adams}, {Addesso}, {Adhikari}, {Adya}, \&
  et~al.}]{2017Natur.551...85A}
{Abbott}, B.~P., {Abbott}, R., {Abbott}, T.~D., {et~al.} 2017, \nat, 551, 85

\bibitem[{Abbott {et~al.}(2018)}]{AbEA2018b}
Abbott, B.~P., {et~al.} 2018, arXiv:1811.12907

\bibitem[{Abbott {et~al.}(2019)}]{AbEA1019}
---. 2019, Phys. Rev., X9, 011001

\bibitem[{{Abbott et al.}(2017{\natexlab{a}})}]{AbEA2017b}
{Abbott et al.} 2017{\natexlab{a}}, Phys. Rev. Lett., 119, 161101.
\newblock \url{https://link.aps.org/doi/10.1103/PhysRevLett.119.161101}

\bibitem[{{Abbott et al.}(2017{\natexlab{b}})}]{AbEA2017e}
---. 2017{\natexlab{b}}, The Astrophysical Journal Letters, 848, L13.
\newblock \url{http://stacks.iop.org/2041-8205/848/i=2/a=L13}

\bibitem[{{Abbott et al.}(2017{\natexlab{c}})}]{AbEA2017f}
---. 2017{\natexlab{c}}, The Astrophysical Journal Letters, 850, L39.
\newblock \url{http://stacks.iop.org/2041-8205/850/i=2/a=L39}

\bibitem[{{Abbott et al.}(2018)}]{AbEA2018}
---. 2018, Living Reviews in Relativity, 21, 3.
\newblock \url{https://doi.org/10.1007/s41114-018-0012-9}

\bibitem[{{Acernese et al}(2015)}]{adVirgo}
{Acernese et al}. 2015, Classical and Quantum Gravity, 32, 024001

\bibitem[{{Alexander} {et~al.}(2017){Alexander}, {Berger}, {Fong}, {Williams},
  {Guidorzi}, {Margutti}, {Metzger}, {Annis}, {Blanchard}, {Brout}, {Brown},
  {Chen}, {Chornock}, {Cowperthwaite}, {Drout}, {Eftekhari}, {Frieman}, {Holz},
  {Nicholl}, {Rest}, {Sako}, {Soares-Santos}, \&
  {Villar}}]{2017ApJ...848L..21A}
{Alexander}, K.~D., {Berger}, E., {Fong}, W., {et~al.} 2017, \apjl, 848, L21

\bibitem[{Andreoni {et~al.}(2019{\natexlab{a}})Andreoni, Goldstein, Anand,
  Coughlin, Singer, Ahumada, Medford, Kool, Webb, Bulla, Bloom, Kasliwal,
  Nugent, Bagdasaryan, Barnes, Cook, Cooke, Duev, Fremling, Gatkine, Golkhou,
  Kong, Mahabal, Mart{\'{\i}}nez-Palomera, Tao, \& Zhang}]{AnGo2019}
Andreoni, I., Goldstein, D.~A., Anand, S., {et~al.} 2019{\natexlab{a}}, The
  Astrophysical Journal, 881, L16.
\newblock \url{https://doi.org/10.3847%2F2041-8213%2Fab3399}

\bibitem[{Andreoni {et~al.}(2019{\natexlab{b}})}]{AnAn2019}
Andreoni, I., {et~al.} 2019{\natexlab{b}}, Publ. Astron. Soc. Pac., 131, 068004

\bibitem[{{Antier} {et~al.}(2019){Antier}, S., V., B., {et~al.}}]{GRANDMApaper}
{Antier}, S., S., A., V., A., B., B., {et~al.} 2019, \mnras

\bibitem[{Arcavi(2018)}]{Arc2018}
Arcavi, I. 2018, Astrophys. J., 855, L23

\bibitem[{Arcavi {et~al.}(2017)Arcavi, McCully, Hosseinzadeh, Howell, Vasylyev,
  Poznanski, Zaltzman, Maoz, Singer, Valenti, Kasen, Barnes, Piran, \& fai
  Fong}]{ArMc2017}
Arcavi, I., McCully, C., Hosseinzadeh, G., {et~al.} 2017, The Astrophysical
  Journal Letters, 848, L33.
\newblock \url{http://stacks.iop.org/2041-8205/848/i=2/a=L33}

\bibitem[{Ascenzi {et~al.}(2019)Ascenzi, Coughlin, Dietrich, Foley,
  Ramirez-Ruiz, Piranomonte, Mockler, Murguia-Berthier, Fryer, Lloyd-Ronning,
  \& Rosswog}]{AsCo2018}
Ascenzi, S., Coughlin, M.~W., Dietrich, T., {et~al.} 2019, Monthly Notices of
  the Royal Astronomical Society, 486, 672.
\newblock \url{https://doi.org/10.1093/mnras/stz891}

\bibitem[{Bauswein {et~al.}(2013)Bauswein, Baumgarte, \& Janka}]{BaBa2013}
Bauswein, A., Baumgarte, T.~W., \& Janka, H.-T. 2013, Phys. Rev. Lett., 111,
  131101.
\newblock \url{https://link.aps.org/doi/10.1103/PhysRevLett.111.131101}

\bibitem[{{Bauswein et al.}(2017)}]{BaJu2017}
{Bauswein et al.} 2017, The Astrophysical Journal Letters, 850, L34.
\newblock \url{http://stacks.iop.org/2041-8205/850/i=2/a=L34}

\bibitem[{Bellm {et~al.}(2018)Bellm, Kulkarni, Graham, Dekany, Smith, Riddle,
  Masci, Helou, Prince, Adams, Barbarino, Barlow, Bauer, Beck, Belicki, Biswas,
  Blagorodnova, Bodewits, Bolin, Brinnel, Brooke, Bue, Bulla, Burruss, Cenko,
  Chang, Connolly, Coughlin, Cromer, Cunningham, De, Delacroix, Desai, Duev,
  Eadie, Farnham, Feeney, Feindt, Flynn, Franckowiak, Frederick, Fremling,
  Gal-Yam, Gezari, Giomi, Goldstein, Golkhou, Goobar, Groom, Hacopians, Hale,
  Henning, Ho, Hover, Howell, Hung, Huppenkothen, Imel, Ip, Ivezi{\'{c}},
  Jackson, Jones, Juric, Kasliwal, Kaspi, Kaye, Kelley, Kowalski, Kramer,
  Kupfer, Landry, Laher, Lee, Lin, Lin, Lunnan, Giomi, Mahabal, Mao, Miller,
  Monkewitz, Murphy, Ngeow, Nordin, Nugent, Ofek, Patterson, Penprase, Porter,
  Rauch, Rebbapragada, Reiley, Rigault, Rodriguez, van Roestel, Rusholme, van
  Santen, Schulze, Shupe, Singer, Soumagnac, Stein, Surace, Sollerman, Szkody,
  Taddia, Terek, Sistine, van Velzen, Vestrand, Walters, Ward, Ye, Yu, Yan, \&
  Zolkower}]{Bellm2018}
Bellm, E.~C., Kulkarni, S.~R., Graham, M.~J., {et~al.} 2018, Publications of
  the Astronomical Society of the Pacific, 131, 018002.
\newblock \url{https://doi.org/10.1088%2F1538-3873%2Faaecbe}

\bibitem[{Blazek {et~al.}(2019)}]{gcn24227}
Blazek, M., {et~al.} 2019, GRB Coordinates Network, 24227

\bibitem[{{Chatterjee et al.}(2019{\natexlab{a}})}]{ChEA2019}
{Chatterjee et al.} 2019{\natexlab{a}}, GRB Coordinates Network, Circular
  Service, No.~24141, \#1 (2019/April-0), 24141

\bibitem[{{Chatterjee et al.}(2019{\natexlab{b}})}]{ChEA2019a}
---. 2019{\natexlab{b}}, GRB Coordinates Network, Circular Service, No.~24237,
  \#1 (2019/April-0), 24237

\bibitem[{{Chornock et al.}(2017)}]{ChBe2017}
{Chornock et al.} 2017, The Astrophysical Journal Letters, 848, L19.
\newblock \url{http://stacks.iop.org/2041-8205/848/i=2/a=L19}

\bibitem[{Connaughton {et~al.}(2016)Connaughton, Burns, Goldstein, Blackburn,
  Briggs, Zhang, Camp, Christensen, Hui, Jenke, Littenberg, McEnery, Racusin,
  Shawhan, Singer, Veitch, Wilson-Hodge, Bhat, Bissaldi, Cleveland,
  Fitzpatrick, Giles, Gibby, von Kienlin, Kippen, McBreen, Mailyan, Meegan,
  Paciesas, Preece, Roberts, Sparke, Stanbro, Toelge, \& Veres}]{CoBu2016}
Connaughton, V., Burns, E., Goldstein, A., {et~al.} 2016, The Astrophysical
  Journal, 826, L6.
\newblock \url{https://doi.org/10.3847%2F2041-8205%2F826%2F1%2Fl6}

\bibitem[{{Cook} {et~al.}(2017){Cook}, {Kasliwal}, {Van Sistine}, {Kaplan},
  {Sutter}, {Kupfer}, {Shupe}, {Laher}, {Masci}, {Dale}, {Sesar}, {Brady},
  {Yan}, \& {Ofek}}]{CoKa2017}
{Cook}, D.~O., {Kasliwal}, M.~M., {Van Sistine}, A., {et~al.} 2017, ArXiv
  e-prints, arXiv:1710.05016

\bibitem[{Cook {et~al.}(2019)}]{gcn24232}
Cook, D.~O., {et~al.} 2019, GRB Coordinates Network, 24232

\bibitem[{Coughlin \& Stubbs(2016)}]{CoSt2016a}
Coughlin, M., \& Stubbs, C. 2016, Experimental Astronomy, 1.
\newblock \url{http://dx.doi.org/10.1007/s10686-016-9503-4}

\bibitem[{Coughlin {et~al.}(2019)Coughlin, Dietrich, Heinzel, Khetan, Antier,
  Christensen, Coulter, \& Foley}]{CoDi2019}
Coughlin, M.~W., Dietrich, T., Heinzel, J., {et~al.} 2019, arXiv:1908.00889

\bibitem[{{Coughlin} {et~al.}(2018){Coughlin}, {Dietrich}, {Margalit}, \&
  {Metzger}}]{CoDi2018b}
{Coughlin}, M.~W., {Dietrich}, T., {Margalit}, B., \& {Metzger}, B.~D. 2018,
  arXiv e-prints, arXiv:1812.04803

\bibitem[{Coughlin {et~al.}(2018{\natexlab{a}})Coughlin, Dietrich, Doctor,
  Kasen, Coughlin, Jerkstrand, Leloudas, McBrien, Metzger, O’Shaughnessy, \&
  Smartt}]{CoDi2018}
Coughlin, M.~W., Dietrich, T., Doctor, Z., {et~al.} 2018{\natexlab{a}}, Monthly
  Notices of the Royal Astronomical Society, 480, 3871.
\newblock \url{http://dx.doi.org/10.1093/mnras/sty2174}

\bibitem[{Coughlin {et~al.}(2018{\natexlab{b}})Coughlin, Tao, Chan, Chatterjee,
  Christensen, Ghosh, Greco, Hu, Kapadia, Rana, Salafia, \& Stubbs}]{CoTo2018}
Coughlin, M.~W., Tao, D., Chan, M.~L., {et~al.} 2018{\natexlab{b}}, Monthly
  Notices of the Royal Astronomical Society, 478, 692.
\newblock \url{http://dx.doi.org/10.1093/mnras/sty1066}

\bibitem[{{Coughlin} {et~al.}(2019){Coughlin}, {Ahumada}, {Anand}, {De},
  {Hankins}, {Kasliwal}, {Singer}, {Bellm}, {Andreoni}, {Cenko}, {Cooke},
  {Copperwheat}, {Dugas}, {Jencson}, {Perley}, {Yu}, {Bhalerao}, {Kumar},
  {Bloom}, {Anupama}, {Ashley}, {Bagdasaryan}, {Biswas}, {Buckley}, {Burdge},
  {Cook}, {Cromer}, {Cunningham}, {D'a{\'\i}}, {Dekany}, {Delacroix},
  {Dichiara}, {Duev}, {Dutta}, {Feeney}, {Frederick}, {Gatkine}, {Ghosh},
  {Goldstein}, {Golkhou}, {Goobar}, {Graham}, {Hanayama}, {Horiuchi}, {Hung},
  {Jha}, {Kong}, {Giomi}, {Kaplan}, {Karambelkar}, {Kowalski}, {Kulkarni},
  {Kupfer}, {La Parola}, {Masci}, {Mazzali}, {Moore}, {Mogotsi}, {Neill},
  {Ngeow}, {Mart{\'\i}nez-Palomera}, {Pavana}, {Ofek}, {Patil}, {Riddle},
  {Rigault}, {Rusholme}, {Serabyn}, {Shupe}, {Sharma}, {Sollerman}, {Soon},
  {Staats}, {Taggart}, {Tan}, {Travouillon}, {Troja}, {Waratkar}, \&
  {Yatsu}}]{CoAh2019b}
{Coughlin}, M.~W., {Ahumada}, T., {Anand}, S., {et~al.} 2019, arXiv e-prints,
  arXiv:1907.12645

\bibitem[{Coughlin {et~al.}(2019{\natexlab{a}})Coughlin, Dekany, Duev, Feeney,
  Kulkarni, Riddle, Ahumada, Burdge, Dugas, Fremling, Hallinan, Prince, \& van
  Roestel}]{Coughlin2018}
Coughlin, M.~W., Dekany, R.~G., Duev, D.~A., {et~al.} 2019{\natexlab{a}},
  Monthly Notices of the Royal Astronomical Society, 485, 1412.
\newblock \url{https://doi.org/10.1093/mnras/stz497}

\bibitem[{Coughlin {et~al.}(2019{\natexlab{b}})Coughlin, Ahumada, Cenko,
  Cunningham, Ghosh, Singer, Bellm, Burns, De, Goldstein, Golkhou, Kaplan,
  Kasliwal, Perley, Sollerman, Bagdasaryan, Dekany, Duev, Feeney, Graham, Hale,
  Kulkarni, Kupfer, Laher, Mahabal, Masci, Miller, Neill, Patterson, Riddle,
  Rusholme, Smith, Tachibana, \& Walters}]{CoAh2019}
Coughlin, M.~W., Ahumada, T., Cenko, S.~B., {et~al.} 2019{\natexlab{b}},
  Publications of the Astronomical Society of the Pacific, 131, 048001

\bibitem[{{Coulter} {et~al.}(2017){Coulter}, {Foley}, {Kilpatrick}, {Drout},
  {Piro}, {Shappee}, {Siebert}, {Simon}, {Ulloa}, {Kasen}, {Madore},
  {Murguia-Berthier}, {Pan}, {Prochaska}, {Ramirez-Ruiz}, {Rest}, \&
  {Rojas-Bravo}}]{2017Sci...358.1556C}
{Coulter}, D.~A., {Foley}, R.~J., {Kilpatrick}, C.~D., {et~al.} 2017, Science,
  358, 1556

\bibitem[{{Cowperthwaite} {et~al.}(2017){Cowperthwaite}, {Berger}, {Villar},
  {et~al.}}]{CoBe2017}
{Cowperthwaite}, P.~S., {Berger}, E., {Villar}, V.~A., {et~al.} 2017, \apjl,
  848, L17

\bibitem[{{D{\'a}lya} {et~al.}(2018){D{\'a}lya}, {Galg{\'o}czi}, {Dobos},
  {Frei}, {Heng}, {Macas}, {Messenger}, {Raffai}, \& {de
  Souza}}]{2018MNRAS.479.2374D}
{D{\'a}lya}, G., {Galg{\'o}czi}, G., {Dobos}, L., {et~al.} 2018, \mnras, 479,
  2374

\bibitem[{{De} {et~al.}(2019){De}, {Adams}, {Kasliwal}, {Coughlin}, {Kasliwal},
  {Hankins}, {Andreoni}, {Anand}, {Singer}, {Ahumada}, {Moore}, {Soon}, Ashley,
  \& {Travouillon}}]{gcn24187}
{De}, K., {Adams}, S.~M., {Kasliwal}, M.~M., {et~al.} 2019, GCN, 24187

\bibitem[{Dekany {et~al.}(2019)Dekany, Smith, {et~al.}}]{DeSm2018}
Dekany, Smith, {et~al.} 2019, Submitted to PASP

\bibitem[{{Drout} {et~al.}(2017){Drout}, {Piro}, {Shappee}, {Kilpatrick},
  {Simon}, {Contreras}, {Coulter}, {Foley}, {Siebert}, {Morrell}, {Boutsia},
  {Di Mille}, {Holoien}, {Kasen}, {Kollmeier}, {Madore}, {Monson},
  {Murguia-Berthier}, {Pan}, {Prochaska}, {Ramirez-Ruiz}, {Rest}, {Adams},
  {Alatalo}, {Ba{\~n}ados}, {Baughman}, {Beers}, {Bernstein}, {Bitsakis},
  {Campillay}, {Hansen}, {Higgs}, {Ji}, {Maravelias}, {Marshall}, {Bidin},
  {Prieto}, {Rasmussen}, {Rojas-Bravo}, {Strom}, {Ulloa},
  {Vargas-Gonz{\'a}lez}, {Wan}, \& {Whitten}}]{2017Sci...358.1570D}
{Drout}, M.~R., {Piro}, A.~L., {Shappee}, B.~J., {et~al.} 2017, Science, 358,
  1570

\bibitem[{{Evans} {et~al.}(2017){Evans}, {Cenko}, {Kennea}, {Emery}, {Kuin},
  {Korobkin}, {Wollaeger}, {Fryer}, {Madsen}, {Harrison}, {Xu}, {Nakar},
  {Hotokezaka}, {Lien}, {Campana}, {Oates}, {Troja}, {Breeveld}, {Marshall},
  {Barthelmy}, {Beardmore}, {Burrows}, {Cusumano}, {Dai}, {D'Avanzo}, {D'Elia},
  {de Pasquale}, {Even}, {Fontes}, {Forster}, {Garcia}, {Giommi},
  {Grefenstette}, {Gronwall}, {Hartmann}, {Heida}, {Hungerford}, {Kasliwal},
  {Krimm}, {Levan}, {Malesani}, {Melandri}, {Miyasaka}, {Nousek}, {O'Brien},
  {Osborne}, {Pagani}, {Page}, {Palmer}, {Perri}, {Pike}, {Racusin}, {Rosswog},
  {Siegel}, {Sakamoto}, {Sbarufatti}, {Tagliaferri}, {Tanvir}, \&
  {Tohuvavohu}}]{2017Sci...358.1565E}
{Evans}, P.~A., {Cenko}, S.~B., {Kennea}, J.~A., {et~al.} 2017, Science, 358,
  1565

\bibitem[{{Flaugher} {et~al.}(2015){Flaugher}, {Diehl}, {Honscheid}, {Abbott},
  {Alvarez}, {Angstadt}, {Annis}, {Antonik}, {Ballester}, {Beaufore},
  {Bernstein}, {Bernstein}, {Bigelow}, {Bonati}, {Boprie}, {Brooks},
  {Buckley-Geer}, {Campa}, {Cardiel-Sas}, {Castand er}, {Castilla}, {Cease},
  {Cela-Ruiz}, {Chappa}, {Chi}, {Cooper}, {da Costa}, {Dede}, {Derylo},
  {DePoy}, {de Vicente}, {Doel}, {Drlica-Wagner}, {Eiting}, {Elliott}, {Emes},
  {Estrada}, {Fausti Neto}, {Finley}, {Flores}, {Frieman}, {Gerdes},
  {Gladders}, {Gregory}, {Gutierrez}, {Hao}, {Holland}, {Holm}, {Huffman},
  {Jackson}, {James}, {Jonas}, {Karcher}, {Karliner}, {Kent}, {Kessler},
  {Kozlovsky}, {Kron}, {Kubik}, {Kuehn}, {Kuhlmann}, {Kuk}, {Lahav}, {Lathrop},
  {Lee}, {Levi}, {Lewis}, {Li}, {Mand richenko}, {Marshall}, {Martinez},
  {Merritt}, {Miquel}, {Mu{\~n}oz}, {Neilsen}, {Nichol}, {Nord}, {Ogando},
  {Olsen}, {Palaio}, {Patton}, {Peoples}, {Plazas}, {Rauch}, {Reil}, {Rheault},
  {Roe}, {Rogers}, {Roodman}, {Sanchez}, {Scarpine}, {Schindler}, {Schmidt},
  {Schmitt}, {Schubnell}, {Schultz}, {Schurter}, {Scott}, {Serrano}, {Shaw},
  {Smith}, {Soares-Santos}, {Stefanik}, {Stuermer}, {Suchyta}, {Sypniewski},
  {Tarle}, {Thaler}, {Tighe}, {Tran}, {Tucker}, {Walker}, {Wang}, {Watson},
  {Weaverdyck}, {Wester}, {Woods}, {Yanny}, \& {DES Collaboration}}]{FlDi2015}
{Flaugher}, B., {Diehl}, H.~T., {Honscheid}, K., {et~al.} 2015, The
  Astronomical Journal, 150, 150

\bibitem[{Ghosh {et~al.}(2017)Ghosh, Chatterjee, Kaplan, Brady, \&
  Sistine}]{GhCh2017}
Ghosh, S., Chatterjee, D., Kaplan, D.~L., Brady, P.~R., \& Sistine, A.~V. 2017,
  Publications of the Astronomical Society of the Pacific, 129, 114503.
\newblock \url{https://doi.org/10.1088%2F1538-3873%2Faa884f}

\bibitem[{{Ghosh et al.}(2019)}]{GhEA2019}
{Ghosh et al.} 2019, GRB Coordinates Network, Circular Service, No.~24377, \#1
  (2019/May-0), 24377

\bibitem[{Goldstein {et~al.}(2017)Goldstein, Veres, Burns, Briggs, Hamburg,
  Kocevski, Wilson-Hodge, Preece, Poolakkil, Roberts, Hui, Connaughton,
  Racusin, von Kienlin, Canton, Christensen, Littenberg, Siellez, Blackburn,
  Broida, Bissaldi, Cleveland, Gibby, Giles, Kippen, McBreen, McEnery, Meegan,
  Paciesas, \& Stanbro}]{GoVe2017}
Goldstein, A., Veres, P., Burns, E., {et~al.} 2017, The Astrophysical Journal,
  848, L14

\bibitem[{Goldstein {et~al.}(2019)}]{GoAn2019}
Goldstein, D.~A., {et~al.} 2019, arXiv:1905.06980

\bibitem[{Graham {et~al.}(2019)Graham, Kulkarni, Bellm, Adams, Barbarino,
  Blagorodnova, Bodewits, Bolin, Brady, Cenko, Chang, Coughlin, De, Eadie,
  Farnham, Feindt, Franckowiak, Fremling, Gezari, Ghosh, Goldstein, Golkhou,
  Goobar, Ho, Huppenkothen, Ivezi{\'{c}}, Jones, Juric, Kaplan, Kasliwal,
  Kelley, Kupfer, Lee, Lin, Lunnan, Mahabal, Miller, Ngeow, Nugent, Ofek,
  Prince, Rauch, van Roestel, Schulze, Singer, Sollerman, Taddia, Yan, Ye, Yu,
  Barlow, Bauer, Beck, Belicki, Biswas, Brinnel, Brooke, Bue, Bulla, Burruss,
  Connolly, Cromer, Cunningham, Dekany, Delacroix, Desai, Duev, Feeney, Flynn,
  Frederick, Gal-Yam, Giomi, Groom, Hacopians, Hale, Helou, Henning, Hover,
  Hillenbrand, Howell, Hung, Imel, Ip, Jackson, Kaspi, Kaye, Kowalski, Kramer,
  Kuhn, Landry, Laher, Mao, Masci, Monkewitz, Murphy, Nordin, Patterson,
  Penprase, Porter, Rebbapragada, Reiley, Riddle, Rigault, Rodriguez, Rusholme,
  van Santen, Shupe, Smith, Soumagnac, Stein, Surace, Szkody, Terek, Sistine,
  van Velzen, Vestrand, Walters, Ward, Zhang, \& Zolkower}]{Graham2018}
Graham, M.~J., Kulkarni, S.~R., Bellm, E.~C., {et~al.} 2019, Publications of
  the Astronomical Society of the Pacific, 131, 078001

\bibitem[{Guessoum {et~al.}(2018)Guessoum, Zitouni, \& Mochkovitch}]{GuZi2019}
Guessoum, N., Zitouni, H., \& Mochkovitch, R. 2018, Astron. Astrophys., 620,
  A131

\bibitem[{{Haggard} {et~al.}(2017){Haggard}, {Nynka}, {Ruan}, {Kalogera},
  {Cenko}, {Evans}, \& {Kennea}}]{2017ApJ...848L..25H}
{Haggard}, D., {Nynka}, M., {Ruan}, J.~J., {et~al.} 2017, \apjl, 848, L25

\bibitem[{{Hallinan} {et~al.}(2017){Hallinan}, {Corsi}, {Mooley}, {Hotokezaka},
  {Nakar}, {Kasliwal}, {Kaplan}, {Frail}, {Myers}, {Murphy}, {De}, {Dobie},
  {Allison}, {Bannister}, {Bhalerao}, {Chandra}, {Clarke}, {Giacintucci}, {Ho},
  {Horesh}, {Kassim}, {Kulkarni}, {Lenc}, {Lockman}, {Lynch}, {Nichols},
  {Nissanke}, {Palliyaguru}, {Peters}, {Piran}, {Rana}, {Sadler}, \&
  {Singer}}]{2017Sci...358.1579H}
{Hallinan}, G., {Corsi}, A., {Mooley}, K.~P., {et~al.} 2017, Science, 358, 1579

\bibitem[{Hotokezaka {et~al.}(2018)Hotokezaka, Nakar, Gottlieb, Nissanke,
  Masuda, Hallinan, Mooley, \& Deller}]{HoNa2018}
Hotokezaka, K., Nakar, E., Gottlieb, O., {et~al.} 2018, arXiv:1806.10596

\bibitem[{Ivezic {et~al.}(2008)Ivezic, Tyson, Allsman, Andrew, \&
  Angel}]{Ivezic2014}
Ivezic, Z., Tyson, J.~A., Allsman, R., Andrew, J., \& Angel, R. 2008,
  arXiv:0805.2366

\bibitem[{Just {et~al.}(2015)Just, Bauswein, Pulpillo, Goriely, \&
  Janka}]{JuBa2015}
Just, O., Bauswein, A., Pulpillo, R.~A., Goriely, S., \& Janka, H.-T. 2015,
  Monthly Notices of the Royal Astronomical Society, 448, 541.
\newblock \url{+ http://dx.doi.org/10.1093/mnras/stv009}

\bibitem[{{Kaiser} {et~al.}(2010){Kaiser}, {Burgett}, {Chambers}, {Denneau},
  {Heasley}, {Jedicke}, {Magnier}, {Morgan}, {Onaka}, \& {Tonry}}]{KaBu2010}
{Kaiser}, N., {Burgett}, W., {Chambers}, K., {et~al.} 2010, in SPIE
  Proceedings, Vol. 7733, Ground-based and Airborne Telescopes III, 77330E

\bibitem[{Kasliwal {et~al.}(2019)Kasliwal, Kasen, Lau, Perley, Rosswog, Ofek,
  Hotokezaka, Chary, Sollerman, Goobar, \& Kaplan}]{KaKa2019}
Kasliwal, M.~M., Kasen, D., Lau, R.~M., {et~al.} 2019, Monthly Notices of the
  Royal Astronomical Society: Letters,
  http://oup.prod.sis.lan/mnrasl/advance-article-pdf/doi/10.1093/mnrasl/slz007/27503647/slz007.pdf.
\newblock \url{https://doi.org/10.1093/mnrasl/slz007}

\bibitem[{{Kasliwal et al.}(2017)}]{KaNa2017}
{Kasliwal et al.} 2017, Science, 358, 1559.
\newblock \url{http://science.sciencemag.org/content/358/6370/1559}

\bibitem[{{Kasliwal et al.}(2019)}]{gcn24191}
---. 2019, GCN, 24191

\bibitem[{{Kilpatrick et al.}(2017)}]{KiFo2017}
{Kilpatrick et al.} 2017, Science, 358, 1583.
\newblock \url{http://science.sciencemag.org/content/358/6370/1583}

\bibitem[{{Klotz} {et~al.}(2008){Klotz}, {Bo{\"e}r}, {Eysseric}, {Damerdji},
  {Laas─Bourez}, {Pollas}, \& {Vachier}}]{2008PASP..120.1298K}
{Klotz}, A., {Bo{\"e}r}, M., {Eysseric}, J., {et~al.} 2008, \pasp, 120, 1298

\bibitem[{{Lipunov} {et~al.}(2010){Lipunov}, {Kornilov}, {Gorbovskoy},
  {Shatskij}, {Kuvshinov}, {Tyurina}, {Belinski}, {Krylov}, {Balanutsa}, \&
  {Chazov}}]{2010AdAst2010E..30L}
{Lipunov}, V., {Kornilov}, V., {Gorbovskoy}, E., {et~al.} 2010, Advances in
  Astronomy, 2010, 349171

\bibitem[{Lyman {et~al.}(2018)Lyman, Lamb, Levan, Mandel, Tanvir, Kobayashi,
  Gompertz, Hjorth, Fruchter, Kangas, Steeghs, Steele, Cano, Copperwheat,
  Evans, Fynbo, Gall, Im, Izzo, Jakobsson, Milvang-Jensen, O'Brien, Osborne,
  Palazzi, Perley, Pian, Rosswog, Rowlinson, Schulze, Stanway, Sutton,
  Th{\"o}ne, de~Ugarte~Postigo, Watson, Wiersema, \& Wijers}]{LyLa2019}
Lyman, J.~D., Lamb, G.~P., Levan, A.~J., {et~al.} 2018, Nature Astronomy, 2,
  751.
\newblock \url{https://doi.org/10.1038/s41550-018-0511-3}

\bibitem[{{Margutti} {et~al.}(2017){Margutti}, {Berger}, {Fong}, {Guidorzi},
  {Alexander}, {Metzger}, {Blanchard}, {Cowperthwaite}, {Chornock},
  {Eftekhari}, {Nicholl}, {Villar}, {Williams}, {Annis}, {Brown}, {Chen},
  {Doctor}, {Frieman}, {Holz}, {Sako}, \&
  {Soares-Santos}}]{2017ApJ...848L..20M}
{Margutti}, R., {Berger}, E., {Fong}, W., {et~al.} 2017, \apjl, 848, L20

\bibitem[{Masci {et~al.}(2018)Masci, Laher, Rusholme, Shupe, Groom, Surace,
  Jackson, Monkewitz, Beck, Flynn, Terek, Landry, Hacopians, Desai, Howell,
  Brooke, Imel, Wachter, Ye, Lin, Cenko, Cunningham, Rebbapragada, Bue, Miller,
  Mahabal, Bellm, Patterson, Juri{\'{c}}, Golkhou, Ofek, Walters, Graham,
  Kasliwal, Dekany, Kupfer, Burdge, Cannella, Barlow, Sistine, Giomi, Fremling,
  Blagorodnova, Levitan, Riddle, Smith, Helou, Prince, \& Kulkarni}]{MaLa2018}
Masci, F.~J., Laher, R.~R., Rusholme, B., {et~al.} 2018, Publications of the
  Astronomical Society of the Pacific, 131, 018003

\bibitem[{{McCully} {et~al.}(2017){McCully}, {Hiramatsu}, {Howell},
  {Hosseinzadeh}, {Arcavi}, {Kasen}, {Barnes}, {Shara}, {Williams},
  {V{\"a}is{\"a}nen}, {Potter}, {Romero-Colmenero}, {Crawford}, {Buckley},
  {Cooke}, {Andreoni}, {Pritchard}, {Mao}, {Gromadzki}, \&
  {Burke}}]{2017ApJ...848L..32M}
{McCully}, C., {Hiramatsu}, D., {Howell}, D.~A., {et~al.} 2017, \apjl, 848, L32

\bibitem[{Metzger(2017)}]{Me2017}
Metzger, B.~D. 2017, Living Rev. Rel., 20, 3

\bibitem[{{Mooley} {et~al.}(2018){Mooley}, {Deller}, {Gottlieb}, {Nakar},
  {Hallinan}, {Bourke}, {Frail}, {Horesh}, {Corsi}, \&
  {Hotokezaka}}]{2018Natur.561..355M}
{Mooley}, K.~P., {Deller}, A.~T., {Gottlieb}, O., {et~al.} 2018, \nat, 561, 355

\bibitem[{{Moore} \& {Kasliwal}(2019)}]{Moore2019}
{Moore}, A.~M., \& {Kasliwal}, M.~M. 2019, Nature Astronomy, 3, 109

\bibitem[{{Nicholl et al.}(2017)}]{NiBe2017}
{Nicholl et al.} 2017, The Astrophysical Journal Letters, 848, L18.
\newblock \url{http://stacks.iop.org/2041-8205/848/i=2/a=L18}

\bibitem[{{O'Brien}(2018)}]{Ob2018}
{O'Brien}, P. 2018, in COSPAR Meeting, Vol.~42, 42nd COSPAR Scientific
  Assembly, E1.15--18--18

\bibitem[{{Pian} {et~al.}(2017){Pian}, {D'Avanzo}, {Benetti}, {Branchesi},
  {Brocato}, {Campana}, {Cappellaro}, {Covino}, {D'Elia}, {Fynbo}, {Getman},
  {Ghirlanda}, {Ghisellini}, {Grado}, {Greco}, {Hjorth}, {Kouveliotou},
  {Levan}, {Limatola}, {Malesani}, {Mazzali}, {Melandri}, {M{\o}ller},
  {Nicastro}, {Palazzi}, {Piranomonte}, {Rossi}, {Salafia}, {Selsing},
  {Stratta}, {Tanaka}, {Tanvir}, {Tomasella}, {Watson}, {Yang}, {Amati},
  {Antonelli}, {Ascenzi}, {Bernardini}, {Bo{\"e}r}, {Bufano}, {Bulgarelli},
  {Capaccioli}, {Casella}, {Castro-Tirado}, {Chassande-Mottin}, {Ciolfi},
  {Copperwheat}, {Dadina}, {De Cesare}, {di Paola}, {Fan}, {Gendre},
  {Giuffrida}, {Giunta}, {Hunt}, {Israel}, {Jin}, {Kasliwal}, {Klose}, {Lisi},
  {Longo}, {Maiorano}, {Mapelli}, {Masetti}, {Nava}, {Patricelli}, {Perley},
  {Pescalli}, {Piran}, {Possenti}, {Pulone}, {Razzano}, {Salvaterra},
  {Schipani}, {Spera}, {Stamerra}, {Stella}, {Tagliaferri}, {Testa}, {Troja},
  {Turatto}, {Vergani}, \& {Vergani}}]{2017Natur.551...67P}
{Pian}, E., {D'Avanzo}, P., {Benetti}, S., {et~al.} 2017, \nat, 551, 67

\bibitem[{Radice {et~al.}(2018)Radice, Perego, Zappa, \& Bernuzzi}]{RaPe2018}
Radice, D., Perego, A., Zappa, F., \& Bernuzzi, S. 2018, The Astrophysical
  Journal Letters, 852, L29.
\newblock \url{http://stacks.iop.org/2041-8205/852/i=2/a=L29}

\bibitem[{Rana {et~al.}(2019)Rana, Anand, \& Bose}]{RaAn2019}
Rana, J., Anand, S., \& Bose, S. 2019, The Astrophysical Journal, 876, 104.
\newblock \url{https://doi.org/10.3847%2F1538-4357%2Fab165a}

\bibitem[{Roberts {et~al.}(2017)Roberts, Lippuner, Duez, Faber, Foucart,
  Lombardi, Ning, Ott, \& Ponce}]{RoLi2017}
Roberts, L.~F., Lippuner, J., Duez, M.~D., {et~al.} 2017, Monthly Notices of
  the Royal Astronomical Society, 464, 3907.
\newblock \url{+ http://dx.doi.org/10.1093/mnras/stw2622}

\bibitem[{Rosswog {et~al.}(2017)Rosswog, Feindt, Korobkin, {et~al.}}]{RoFe2017}
Rosswog, S., Feindt, U., Korobkin, O., {et~al.} 2017, Class. Quant. Grav., 34,
  104001

\bibitem[{{Shappee} {et~al.}(2014){Shappee}, {Prieto}, {Grupe}, {Kochanek},
  {Stanek}, {De Rosa}, {Mathur}, {Zu}, {Peterson}, {Pogge}, {Komossa}, {Im},
  {Jencson}, {Holoien}, {Basu}, {Beacom}, {Szczygie{\l}}, {Brimacombe},
  {Adams}, {Campillay}, {Choi}, {Contreras}, {Dietrich}, {Dubberley},
  {Elphick}, {Foale}, {Giustini}, {Gonzalez}, {Hawkins}, {Howell}, {Hsiao},
  {Koss}, {Leighly}, {Morrell}, {Mudd}, {Mullins}, {Nugent}, {Parrent},
  {Phillips}, {Pojmanski}, {Rosing}, {Ross}, {Sand}, {Terndrup}, {Valenti},
  {Walker}, \& {Yoon}}]{ShPr2014}
{Shappee}, B.~J., {Prieto}, J.~L., {Grupe}, D., {et~al.} 2014, The
  Astrophysical Journal, 788, 48

\bibitem[{{Shappee} {et~al.}(2017){Shappee}, {Simon}, {Drout}, {Piro},
  {Morrell}, {Prieto}, {Kasen}, {Holoien}, {Kollmeier}, {Kelson}, {Coulter},
  {Foley}, {Kilpatrick}, {Siebert}, {Madore}, {Murguia-Berthier}, {Pan},
  {Prochaska}, {Ramirez-Ruiz}, {Rest}, {Adams}, {Alatalo}, {Ba{\~n}ados},
  {Baughman}, {Bernstein}, {Bitsakis}, {Boutsia}, {Bravo}, {Di Mille}, {Higgs},
  {Ji}, {Maravelias}, {Marshall}, {Placco}, {Prieto}, \&
  {Wan}}]{2017Sci...358.1574S}
{Shappee}, B.~J., {Simon}, J.~D., {Drout}, M.~R., {et~al.} 2017, Science, 358,
  1574

\bibitem[{{Shawhan et al.}(2019)}]{ShEA2019}
{Shawhan et al.} 2019, GRB Coordinates Network, Circular Service, No.~24098,
  \#1 (2019/April-0), 24098

\bibitem[{Singer \& Price(2016)}]{SiPr2016}
Singer, L.~P., \& Price, L.~R. 2016, Phys. Rev., D93, 024013

\bibitem[{Singer {et~al.}(2014)Singer, Price, Farr, {et~al.}}]{SiPr2014}
Singer, L.~P., Price, L.~R., Farr, B., {et~al.} 2014, Astrophys. J., 795, 105

\bibitem[{{Singer et al.}(2013)}]{SiCe2013}
{Singer et al.} 2013, The Astrophysical Journal Letters, 776, L34.
\newblock \url{http://stacks.iop.org/2041-8205/776/i=2/a=L34}

\bibitem[{{Singer et al.}(2019{\natexlab{a}})}]{SiEA2019}
---. 2019{\natexlab{a}}, GRB Coordinates Network, Circular Service, No.~24069,
  \#1 (2019/April-0), 24069

\bibitem[{{Singer et al.}(2019{\natexlab{b}})}]{SiEA2019a}
---. 2019{\natexlab{b}}, GRB Coordinates Network, Circular Service, No.~24168,
  \#1 (2019/April-0), 24168

\bibitem[{{Singer et al.}(2019{\natexlab{c}})}]{SiEA2019b}
---. 2019{\natexlab{c}}, GRB Coordinates Network, Circular Service, No.~24228,
  \#1 (2019/May-0), 24228

\bibitem[{{Smartt et al.}(2017)}]{SmCh2017}
{Smartt et al.} 2017, Nature, 551, 75 EP .
\newblock \url{http://dx.doi.org/10.1038/nature24303}

\bibitem[{Tonry {et~al.}(2018)Tonry, Denneau, Heinze, Stalder, Smith, Smartt,
  Stubbs, Weiland, \& Rest}]{ToDe2018}
Tonry, J.~L., Denneau, L., Heinze, A.~N., {et~al.} 2018, Publications of the
  Astronomical Society of the Pacific, 130, 064505.
\newblock \url{http://stacks.iop.org/1538-3873/130/i=988/a=064505}

\bibitem[{{Troja} {et~al.}(2017){Troja}, {Piro}, {van Eerten}, {Wollaeger},
  {Im}, {Fox}, {Butler}, {Cenko}, {Sakamoto}, {Fryer}, {Ricci}, {Lien}, {Ryan},
  {Korobkin}, {Lee}, {Burgess}, {Lee}, {Watson}, {Choi}, {Covino}, {D'Avanzo},
  {Fontes}, {Gonz{\'a}lez}, {Khandrika}, {Kim}, {Kim}, {Lee}, {Lee}, {Kutyrev},
  {Lim}, {S{\'a}nchez-Ram{\'{\i}}rez}, {Veilleux}, {Wieringa}, \&
  {Yoon}}]{2017Natur.551...71T}
{Troja}, E., {Piro}, L., {van Eerten}, H., {et~al.} 2017, \nat, 551, 71

\bibitem[{{Utsumi} {et~al.}(2017){Utsumi}, {Tanaka}, {Tominaga}, {Yoshida},
  {Barway}, {Nagayama}, {Zenko}, {Aoki}, {Fujiyoshi}, {Furusawa}, {Kawabata},
  {Koshida}, {Lee}, {Morokuma}, {Motohara}, {Nakata}, {Ohsawa}, {Ohta},
  {Okita}, {Tajitsu}, {Tanaka}, {Terai}, {Yasuda}, {Abe}, {Asakura}, {Bond},
  {Miyazaki}, {Sumi}, {Tristram}, {Honda}, {Itoh}, {Itoh}, {Kawabata},
  {Morihana}, {Nagashima}, {Nakaoka}, {Ohshima}, {Takahashi}, {Takayama},
  {Aoki}, {Baar}, {Doi}, {Finet}, {Kanda}, {Kawai}, {Kim}, {Kuroda}, {Liu},
  {Matsubayashi}, {Murata}, {Nagai}, {Saito}, {Saito}, {Sako}, {Sekiguchi},
  {Tamura}, {Tanaka}, {Uemura}, \& {Yamaguchi}}]{2017PASJ...69..101U}
{Utsumi}, Y., {Tanaka}, M., {Tominaga}, N., {et~al.} 2017, \pasj, 69, 101

\bibitem[{Veitch {et~al.}(2015)}]{VeRa2015}
Veitch, J., {et~al.} 2015, Phys. Rev., D91, 042003

\bibitem[{Veres {et~al.}(2019)Veres, dal Canton, Burns, Goldstein, Littenberg,
  Christensen, Preece, \& D.}]{VeDa2019}
Veres, P., dal Canton, T., Burns, E., {et~al.} 2019, arXiv:1905.08755

\bibitem[{{Waratkar} {et~al.}(2019){Waratkar}, {Kumar}, {Bhalerao},
  {Karambelkar}, {Anupama}, \& {Stanzin}}]{gcn24316}
{Waratkar}, G., {Kumar}, H., {Bhalerao}, V., {et~al.} 2019, GCN, 24316

\bibitem[{Wu {et~al.}(2016)Wu, Fern{\'a}ndez, Mart??nez-Pinedo, \&
  Metzger}]{WuFe2016}
Wu, M.-R., Fern{\'a}ndez, R., Mart??nez-Pinedo, G., \& Metzger, B.~D. 2016,
  Monthly Notices of the Royal Astronomical Society, 463, 2323.
\newblock \url{+ http://dx.doi.org/10.1093/mnras/stw2156}

\end{thebibliography}



\end{document}